 \documentclass[twocol]{ametsocV6.1}

\title{Model Training, Data Assimilation, and Forecast Experiments  with a Hybrid Atmospheric Model that Incorporates Machine Learning}

\authors{Dylan Elliott,\aff{a}
Troy Arcomano,\aff{b} 
Istvan Szunyogh,\aff{a}\correspondingauthor{Istvan Szunyogh, szunyogh@tamu.edu} 
Brian R. Hunt,\aff{c}  
}

\affiliation{\aff{a}{Texas A\&M University}\\
\aff{b}{Argonne National Laboratory}\\
\aff{c}{University of Maryland}
}

\abstract{The hybrid model combines the physics-based primitive-equations model SPEEDY with a machine learning-based (ML-based) model component, while ERA5 reanalyses provide the presumed true states of the atmosphere. Six-hourly simulated noisy observations are generated for a 30-year ML training period and a one-year testing period. These observations are assimilated with a Local Ensemble Transform Kalman Filter (LETKF), and a 10-day deterministic forecast is also started from each ensemble mean analysis of the testing period. In the first experiment, the physics-based model provides the background ensemble members and the 10-day deterministic forecasts. In the other three experiments, the hybrid model plays the same role as the physics-based model in the first experiment, but it is trained on a different data set in each experiment. These training data sets are analyses obtained by using the physics-based model (second experiment), the hybrid model of the previous experiment (third experiment), and for comparison, ERA5 reanalyses (fourth experiment). The results of the experiments show that hybridizing the model can substantially improve the accuracy of the analyses and forecasts. When the model is trained on ERA5 reanalyses, the biases of the analyses are negligible and the magnitude of the flow-dependent part of the analysis errors is greatly reduced. While the gains in analysis accuracy are distinctly more modest in the other two hybrid model experiments, the gains in forecast accuracy tend to be larger in those experiments after 1-3 forecast days. However, these extra gains of forecast accuracy are achieved, in part, by a modest gradual reduction of the spatial variability  of the forecasts.} 

\begin{document}

\maketitle

%
%
%
\statement
This is the first study to investigate the analysis and forecast effects of the interactions between ML model training and data assimilation for a realistic hybrid model of the atmospheric dynamics based on the primitive equations.
%
%

%
\section{Introduction}
\begin{figure}[h]
\centerline{\includegraphics[width=0.5\textwidth]{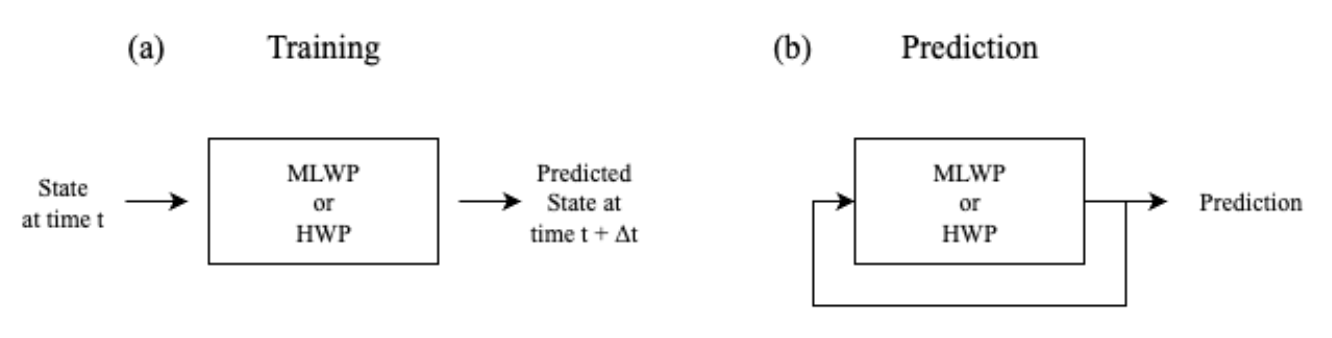}}
\caption{Schematic illustration of the relationship between training and time marching for an MLWP or HWP model. (a) The model is trained to make $\Delta t$ long forecasts. (b) When the model is used in prediction mode, longer term forecasts are prepared by time-marching the learnt mapping of the state. Preparing a  $n \Delta$ lead time forecast requires $n$ iterations ``time steps''.  }
\label{fig:0}
\end{figure}

\emph{Machine learning-based weather prediction} (MLWP) models \citep[e.g.][]{Arcomano2020,Weyn2021,Pathak2022,Lam2023,Bi2023} and \emph{hybrid weather prediction} (HWP) models that incorporate \emph{machine learning} (ML) \citep[e.g.,][]{Arcomano2022,Arcomano2023,Kochkov2024} are typically trained on decades long time series of reanalysis data. These models are trained by \emph{supervised learning}: the models learn to predict the reanalysis at time $t+\Delta t$ based on the reanalysis at time $t$, and in some models also at time $t-\Delta t$. The  length of a ``time step"  usually varies between $\Delta t =1$\,h and $\Delta t = 24$\,h depending on the model and the intended length of the forecasts, which can be obtained by time-marching the ``time steps'' as in a conventional numerical weather prediction (NWP) model (Fig.~\ref{fig:0}).  Another similarity to an NWP model forecast is that a MLWP or HWP model forecast is also started from a real-time analysis of the atmospheric state.

Training on reanalyses is often referred to as \emph{offline training} \citep{Boucquet2021}, because the observations-based estimates of the atmospheric states (reanalyses) used for training are obtained by a data assimilation (DA) process that is independent of the training process. In this setting, the models are not trained directly on observations, and the training process has no information about the errors of the analyses. The alternative to offline training is \emph{online training} \citep{Boucquet2021,Malartic2022,Farchi2021,Farchi2023}. Online training takes advantage of the fact that DA and ML model training both use observational information to solve an estimation problem: while the primary goal of DA is the estimation of the atmospheric state, the goal of ML model training is the estimation of the trainable parameters of the ML model. Offline training separates these two estimation problems  by using a time series of retrospective estimates (observational reanalyses) of the atmospheric states for training. Online training, in contrast, estimates the state and the trainable ML model parameters together in a sequential DA process: at a specific analysis time, the latest observations are assimilated to update a background state, which is a prior estimate of the atmospheric state, and background ML model parameters, which are prior estimates of the trainable ML model parameters. These backgrounds represent the knowledge about the state and parameters from the observations assimilated in the past.

An operational DA system uses an operational NWP model to obtain the background state from the previous completed analysis. The accuracy of the new analysis critically depends on the ability of the forecast model to provide an accurate background. In addition, in a modern DA system, which uses a 4D-Var or ensemble-based Kalman filter scheme, the model is also used for the prediction of the probability distribution of the uncertainty in the background state. This information about the uncertainty in the knowledge of the state is used by the analysis scheme to perform a statistical interpolation by the proper weighting of the background state and the inherently noisy observations. The quality of the prediction of the uncertainty in the background state depends on the ability of the model to capture the unstable, neutral, and stable directions of the state space along the state space trajectory.

It has long been known from filtering theory that the accuracy of the final state estimate (the analysis in meteorological terminology) can be improved by estimating the effect of the forecast model errors on the background, and then correcting for that in the calculation of a state estimate  \citep{Friedland1969}. The online estimation of the effect of model errors can be done by the augmentation of the state vector with the components of the correction term, or the parameters of a parameterized correction term, and adding an evolution equation for the added components of the augmented state \citep{Jazwinski1970}. The two challenging aspects of this approach are to find a proper evolution equation for the correction terms and to keep the increased dimensionality of the estimation problem computationally manageable. Addressing these challenges usually requires making further assumptions. For instance, the operational DA system of the European Centre for Medium-Range Weather Forecasts (ECMWF) uses a formulation of weak-constraint 4D-Var \citep{Tremolet2006,Laloyaux2020} that assumes that the model error correction term is an additive correction to the model state in an atmospheric column, its components are independent, and it is constant throughout the assimilation time window (the time interval from which observations are assimilated at an analysis time). \citet{Farchi2021} developed an ML version of this algorithm in which the correction term is modeled by a neural network. In their algorithm, the trainable parameters of the neural network rather than the components of the additive model error correction term are assumed to be constant throughout the assimilation time window. In this approach, the HWP model is formed by adding the ML-based correction term to the physics-based NWP model forecast.  \citet{Farchi2023} developed a simplified and computationally more efficient version of this algorithms for the incremental formulation of 4D-Var. They demonstrated the potential of the approach by carrying out simulated observations experiments with a two-level quasi-geostrophic model. They showed that their online training procedure led to more accurate analyses and forecasts than offline training.

We present the results of our first attempt to use the hybrid model of \citet{Arcomano2022} for DA. The hybridization strategy of this model \citep{Pathak2018,Wikner2020} differs from that of \citet{Farchi2023} in several respects, of which we highlight only the most important ones. First, the trainable parameters are the entries of two (non-diagonal) weight matrices that determine the optimal combination of the physics-based and data-driven (reservoir-based) description of the evolving atmospheric state rather than the parameters of an additive correction term. Second, the model can learn about the relationships between the state variables, not only at the different vertical levels in an atmospheric column, but also at the different horizontal locations within a local neighborhood. In contrast, the trainable parameters of  \cite{Farchi2023} are global parameters that describe the errors for a  vertical column of the model atmosphere. Third, it uses an ML architecture based on \emph{reservoir computing} (RC) \citep{Jaeger2001,Lukosevicius2009,Lukosevicius2012} rather than multiple dense neural layers. The price to be paid for the added flexibility of the hybridization approach of \citet{Arcomano2022} is the substantially larger number of ML model parameters that must be trained. In order to assess the potential advantages and disadvantages of training a HWP model online, directly from observations, with a global circulation model, we follow the iterative approach of \cite{Wikner2021}; see also \cite{Brajard2020} for a similar approach with MLWP.  This approach alternates data assimilation with offline training of the model, with the goal of converging toward a model and a time series of analyses that optimize both the parameter and state estimates, as in online training.

The structure of the paper is as follows. Section~\ref{sec:methodology} provides brief descriptions of the hybrid model  and DA system used in our analysis-forecast experiments. It also explains the rationale for the specific design of the experiments. Section~\ref{sec:results} presents the results of the experiments, while Section~\ref{sec:conclusions} offers our conclusions and outlines the   plans for the next steps of our research into the integration of ML model training and DA. 

\section{Methodology} \label{sec:methodology}

\subsection{The hybrid model}
All analysis-forecast experiments of this study are carried out with the version of the hybrid model described in \citet{Arcomano2022}. Later versions of the model \citep{Arcomano2023,Patel2023} added ML-based capabilities for the prediction of precipitation and sea-surface-temperature, but these capabilities are not used in this study. 

\paragraph{Physics-based model component} The physics-based component of the model is the Simplified Parameterization, primitive-Equation Dynamics model (SPEEDY) \citep{Molteni2003,Kucharski2006}. Though SPEEDY is a low-resolution model, which was developed for academic research rather than operational numerical weather prediction, it can provide skillful global numerical predictions of large- and synoptic-scale atmospheric motions for several days. It uses the spectral transform technique to solve the atmospheric primitive equations at resolution T30. Model input and output are provided on the corresponding latitude-longitude grid, which has 48 grid points in the meridional direction and 96 grid points in the zonal direction. This grid provides a $3.75^\circ \times 3.75^\circ$ horizontal resolution that corresponds to about a 300\,km grid spacing in the mid-latitudes. The model has eight vertical pressure $\sigma$-levels (0.025, 0.095, 0.20, 0.34, 0.51, 0.685, 0.835, and 0.95), where $\sigma$ is the ratio of pressure to the surface pressure. Though the top layers of the model are in the lower stratosphere, their purpose is to soften the artificial effects of not having higher atmospheric levels to realistically handle vertically propagating waves, rather than to capture lower stratospheric dynamics. The prognostic variables of the model are the two horizontal components of the wind, temperature, and specific humidity at the 8 sigma levels and surface pressure. 

\paragraph{Localization strategy} The hybrid model uses the model grid of SPEEDY for the representation of the global atmospheric state. Thus, the format of the input and output data is the same for the two models. We introduce the notation $\mathbf{v}(t)$ for the vector that represents the global atmospheric state on this common grid at time $t$. For the hybrid calculations, the global grid is broken up horizontally into 1152 disjoint local domains (volumes) such that each local domain includes $2 \times 2 \times 8=32$ grid points: 2 grid points in each horizontal direction and all 8 model levels in the vertical direction. The local states in the local domains are represented by local state vectors $\mathbf{v}_\ell(t)$, $\ell=1,2,\dots,1152$. These local state vectors are formed by the local components of $\mathbf{v}(t)$ after a location-dependent standardization that makes the components non-dimensional. The purpose of the standardization is to ensure that the different state variables vary in the same range (see \citet{Arcomano2022} for details). The dimension of a local state vector is $m = 4 \times 32 + 4 = 132$: 4 state variables are defined at the 32 grid points of the local domain, while the surface pressure is represented by 4 horizontal grid points.

\paragraph{Calculations of a ``time step"} After training, the hybrid model calculations of a $\Delta t = 6~h$  long ``time step'' to obtain $\mathbf{v}^h(t+\Delta t)$ start from the global hybrid model state $\mathbf{v}^h(t)$. First, the (physics-based) global SPEEDY forecast $\mathbf{v}^p (t + \Delta t)$ is computed using $\mathbf{v}^h(t)$ as the initial condition. (The length of a numerical time step used in SPEEDY for this forecast is $\Delta t^\prime = 0.25$\,h, so that $\Delta t/\Delta t^\prime = 24$.) Then, the local hybrid model solutions are computed for the local domains in parallel by 
\begin{equation} \label{eq:hybrid}
\mathbf{v}_\ell^h (t + \Delta t) = \mathbf{W}^p_\ell \mathbf{v}^p_\ell (t + \Delta t) + \mathbf{W}^r_\ell \mathbf{r}_\ell (t + \Delta t), \  \  \  \ \ell=1,2,\dots,1152,
\end{equation}
where $\mathbf{v}^p_\ell (t + \Delta t)$ is formed of the relevant standardized components of $\mathbf{v}^p (t + \Delta t)$, while $\mathbf{r}_\ell (t + \Delta t)$ represents the state of the local reservoir, which is a high-dimensional dynamical system described in the next paragraph. The entries of the $m \times m$ weight matrix $\mathbf{W}_\ell^p$ and $m \times D_r$ weight matrix $\mathbf{W}^r_\ell$, where  $D_r=6000$ is the dimension of the reservoir, are the trainable parameters of the hybrid model. Converting the standardized components of $\mathbf{v}_\ell^h (t + \Delta t)$ back to dimensional physical quantities and concatenating the resulting dimensional local state vectors to obtain $\mathbf{v}^h(t+\Delta t)$ completes the ``time step''.

\paragraph{Reservoir dynamics} The evolution equation of a local reservoir is 
\begin{equation} \label{eq:resdyn}
{\bf r}_\ell (t+\Delta t) = \tanh{\left[\mathbf{A}_\ell \mathbf{r}_\ell (t) + \mathbf{B}_\ell \mathbf{u}_\ell^h(t) \right]}, \  \  \  \ \ell=1,2,\dots,1152.
\end{equation}
The input vector $\mathbf{u}^h_\ell(t)$ is formed like $\mathbf{v}_\ell^h(t)$, except that it represent the state in an extended local domain. Compared to the corresponding local domain, an extended local domain includes an extra column of grid points on both sides in the zonal direction and an extra row of grid points on both sides in the meridional direction. Hence, the extended local regions have $4 \times 4$ rather than $2 \times 2$ horizontal grid points, such that the neighboring horizontal regions overlap by one grid point on each side. This overlap ensures that information about the atmospheric state can flow between reservoirs of neighboring local regions. The dimension of the extended local state vectors is $n= 4 \times 16 \times 8 + 16 =528$: there are 16 horizontal grid points at each of the 8 vertical levels and the surface pressure is represented by 16 horizontal grid points. Matrix ${\bf A}_\ell$ is a sparse $D_r \times D_r$ random matrix, while  matrix $\mathbf{B}_\ell$  is a sparse $D_r \times n$ random matrix. The parameters that control the statistical properties of the random entries of these matrices are \emph{hyperparameters} of the hybrid model: parameters whose value is determined by experimentation (``model tuning'') rather than model training. Specifically, each random entry of $\mathbf{A}$ is generated with probability $\kappa/D_r$ of not being zero, where $\kappa=6$, the entries of ${\bf A}$ are scaled such that the largest eigenvalue of ${\bf A}$ varies depending on the geographical latitude between $\rho=0.3$ and $\rho=0.7$, and the entries of $\mathbf{B}_\ell$ are chosen from a uniform distribution on the interval $(-0.5,0.5)$. (The dimensions $D_r$, $m$, and $n$ are other examples of hyperparameters.) The vector-to-vector activation function $\tanh{(\cdot)}$ and its argument  both have $D_r$ components: each component of $\tanh{(\cdot)}$ is the hyperbolic tangent of the corresponding component of the argument. 

\paragraph{Training} Training data consists of a time series of global analysis states $\mathbf{v}^a (k \Delta t)$ for $k=0,1,\dots,K$. For $k=0,1,\dots,K-1$, SPEEDY forecasts $\mathbf{v}^p (k \Delta t+\Delta t)$ are computed from initial conditions $\mathbf{v}^a (k \Delta t)$, and reservoir states ${\bf r}_\ell(k\Delta t + \Delta t)$ are computed using Eq.\,(\ref{eq:resdyn}) with  $\mathbf{u}_\ell^h$ replaced by  $\mathbf{u}_\ell^a$, where $\mathbf{u}_\ell^a$ is formed from $\mathbf{u}_\ell^a$ in the same way as $\mathbf{u}_\ell^h$ from $\mathbf{u}^h$, except that each component of ${\bf u}^a_\ell$ is multiplied by $1 + \delta$, where $\delta$ is a zero-mean, normally distributed random number chosen independently for each component and at each time step. The addition of such noise has been found beneficial for the stability of RC-based ML models even in controlled experiments, in which the model is trained on error-free observations of the modeled system. The usual explanation for this behavior of the models is that training on noisy data can help them learn to return to the attractor of the dynamics in the presence of noise that is expected to arise in future forecasts \citep[e.g.,][]{Jaeger2001,Lukosevicius2009,Lukosevicius2012,Wikner2024}. The standard deviation of the multiplicative noise factor $1+\delta$ is a hyperparameter of the hybrid model that has a value of $0.2$ in our experiments.

Offline training is carried out by seeking the entries of matrix $\mathbf{W}_\ell = \left( \mathbf{W}^p_\ell \ \ \mathbf{W}^r_\ell \right)$ that minimize the cost function 
\begin{flalign} \label{eq:costf}
& J_\ell (\boldsymbol{W}_\ell)  =  \sum_{k=0}^{K-1}  \| \mathbf{v}^h_\ell \left(k \Delta t+\Delta t, \mathbf{W}_\ell \right) - \mathbf{v}^a_\ell (k \Delta t +\Delta t) \|^2 &  \\
&  +   \beta^p \| \mathbf{W}^p_\ell \|_F^2 + \beta^r \| \mathbf{W}^r_\ell \|_F^2,  \  \   \ \ell=1,2,\dots,1152, & \nonumber
\end{flalign}
where $\mathbf{v}^h_\ell \left(k \Delta t+\Delta t, \mathbf{W}_\ell \right)$ is computed according to Eq.\,(\ref{eq:hybrid}).
In our experiments, $k=0$ corresponds to 0000\,UTC January 1, 1981 and $k=K-1$ to 1800\,UTC December 31, 2009. 
The first term of Eq.\,(\ref{eq:costf}) quantifies how well the model fits the training data. The last two terms of the cost function are regularization terms whose role is to prevent over-fitting to the training data in tandem with the added noise \citep{Tikhonov+Arsenin1977}. The symbol $\| \cdot \|_F$ stands for the Frobenius matrix norm, which is defined such that $\| \cdot \|^2_F$ is equal to the sum of the squares of the entries of its matrix argument. The values of the two regularization parameters, which are also hyperparameters, are $\beta^r=10^{-4}$ and $\beta^p=1$ in our experiments. 
The minimization problem for $J_\ell (\boldsymbol{W}_\ell)$ can be solved directly (analytically) without the use of a numerical minimization algorithm and its solution is a ridge regression. 

\paragraph{The role of the physics- and RC-based model component}
If the weight matrix of the physics-based component is set to $\mathbf{W}^p_\ell={\bf 0}$, the hybrid model becomes a MLWP model. If the weight matrix of the reservoir component is set to $\mathbf{W}^r_\ell={\bf 0}$, the model learns to perform a linear regression of the $\Delta t$-long physics-based local forecasts to better fit the training data. Both of these configurations of the model have been found to have considerable forecast skill \citep{Arcomano2020,Arcomano2022}: the model with  $\mathbf{W}^p_\ell={\bf 0}$ provide more accurate global forecasts than SPEEDY up to 3 days for the temperature and up to 5 days for the specific humidity, and with $\mathbf{W}^r_\ell={\bf 0}$  for all variables up to 5 days. In fact, in this forecast range, the latter model performs almost as well as the full hybrid model, except for the temperature. Beyond this range, however, this version of the model starts to exhibit unrealistic behavior and rapidly becomes unstable. In contrast, the full hybrid model remains stable and  maintains a realistic climate at the limited resolution of the model for several decades (the longest period tested has been 70 years). These results suggest that the role of the two weight matrices is more than just to determine the optimal weighting of the two model components:  $\mathbf{W}^p$ also performs a linear transformation of the physics-based forecast, while $\mathbf{W}^r$ also reads out the prognostic state variables from their high-dimensional randomized representation by the reservoir. It should be noted that Eq.~(\ref{eq:hybrid}) could be used for an ML-based additive model correction by making the choice $\mathbf{W}^p={\bf I}$, but such a configuration of the model has not been tested, yet.

\subsection{The observations}
While the true state space trajectory of the atmosphere is not known, in our controlled experiments, we assume ERA5 reanalyses \citep{Herschbach2020} represent such a trajectory. Obtaining the ``true'' states on the model grid of SPEEDY requires a spatial interpolation of the ERA5 reanalyses. We start the interpolation from the ERA5 reanalyses of the prognostic state variables of SPEEDY at constant pressure surfaces. These fields and the ERA5 surface pressure reanalyses are first interpolated horizontally with a 2-dimensional quadratic B-spline interpolation to the horizontal locations of the grid points of SPEEDY. Then, the sigma values associated with the constant pressure levels are computed for each horizontal grid point location.  Finally, the values of the prognostic variables are interpolated with a 1-dimensional cubic B-spline to the constant $\sigma$ surfaces of SPEEDY. In this procedure, we do not adjust the interpolated surface pressure values to the low-resolution model orography of SPEEDY. This way the experiment design mimics the real-life situation that the surface pressure associated with the reduced resolution model orography is different from the surface pressure associated with the actual orography.

\begin{figure}[h]
\centerline{\includegraphics[width=0.5\textwidth]{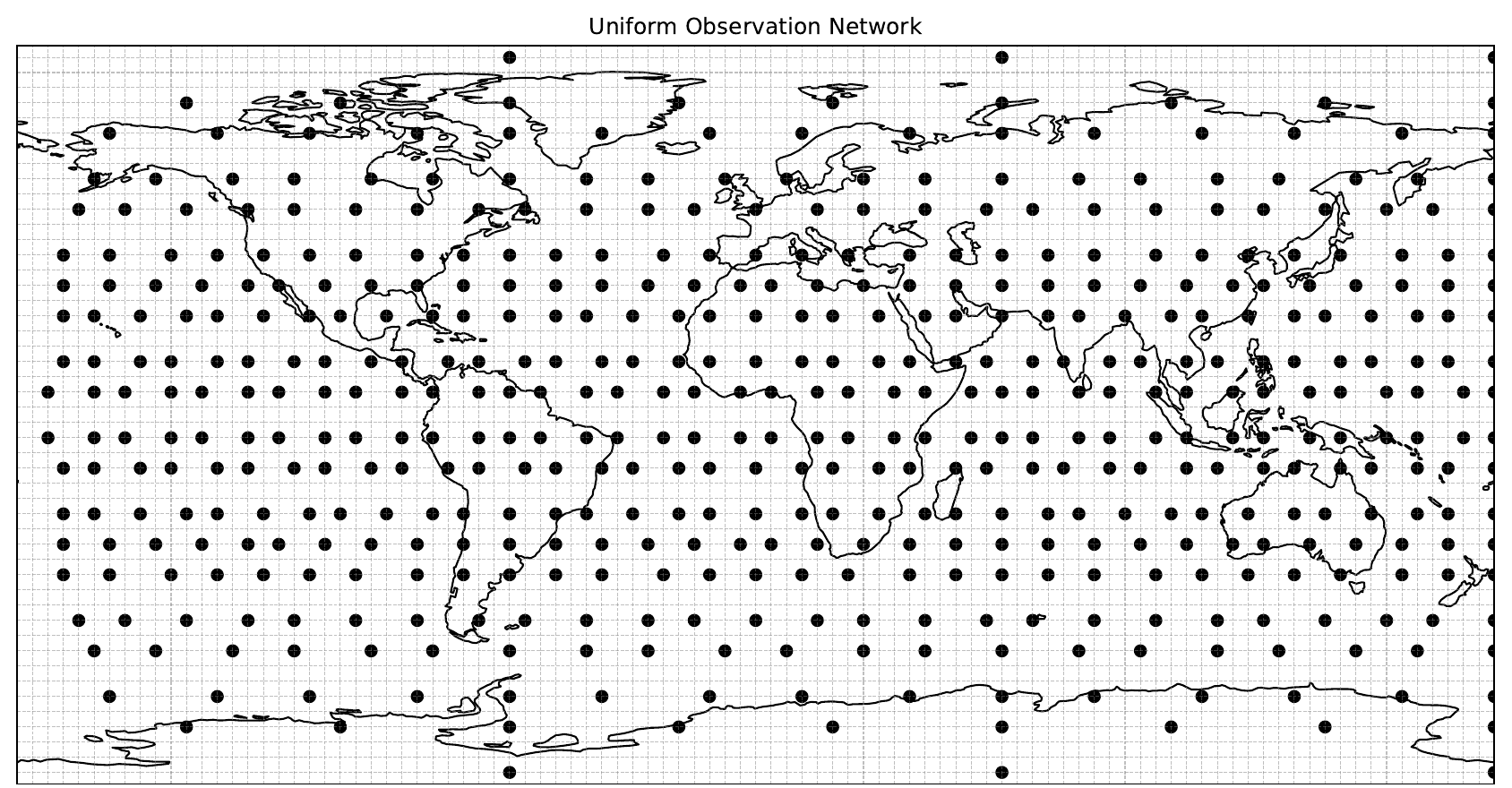}}
\caption{The simulated observing network. This network consists of the same 500 observed horizontal grid points at each model level (about 11\% of the horizontal grid points per vertical level of the model). The dots mark the locations of the observations.}
\label{fig:1}
\end{figure}

We generate simulated observations by adding random observation noise to the ``true'' states at each analysis time. The locations of the simulated observations do not change in time and they always fall on model grid points. The horizontal locations of the grid points are selected such that they provide a near uniform horizontal areal coverage (Fig.~\ref{fig:1}). All prognostic variables are observed at each model level at the selected horizontal locations. The randomly generated observation noise has a normal distribution with mean zero and a prescribed standard deviation, which is 1\,m/s for the two horizontal components of the wind, 1\,K for the temperature, 1\,g/kg for the specific humidity, and 1\,hPa for the surface pressure.

\subsection{The DA scheme}
We use the local ensemble transform Kalman filter (LETKF) scheme for DA. This scheme was developed in a series of paper by \citet{Ott2004,Hunt2007,Szunyogh2008}, and it became one of the most widely used DA  schemes for spatiotemporally chaotic systems, including the atmosphere. It has also been included \citep[e.g.,][]{Park2023} in the Joint Effort for Data assimilation Integration (JEDI) system, which is a community effort for DA code integration led by the Joint Center for Satellite Data Assimilation (JCSDA), a partnership between NOAA, NASA, the US Navy and US Air Force. We use a computer code of the LETKF that was originally developed by Takemasa Miyoshi and was made publicly available with some modifications by \citet{Hatfield2018}. Because this code was developed for the assimilation of simulated observations based on a state space trajectory of SPEEDY, we made some minor modifications to the code to accommodate the assimilation of the ERA5-based simulated observations. 

As all other ensemble-based DA schemes, the LETKF uses an ensemble of model forecasts for the prediction of the spatiotemporal  evolution of the background probability distribution, which it assumes to be a multivariate normal distribution. Such a distribution can be described by the mean (a vector) and the covariance matrix of the distribution. The ensemble mean forecast is the prediction of the mean, while the ensemble perturbations, which are defined by the difference between the ensemble members and the ensemble mean, yield the prediction of the covariance matrix. The analysis is obtained by making a correction to the background (the ensemble mean) in the linear (vector) space spanned by the ensemble perturbations. The accuracy of an analysis strongly depends on the quality of the forecast model, because it affects both the accuracy of the background and the effectiveness of the ensemble perturbations in capturing the space of uncertainties in the knowledge of the state \citep{Szunyogh2005,Kuhl2007}. A more accurate background is a better staring point for the analysis, especially for the unobserved state variables, while a better prediction of the space of uncertainties allows the scheme to make a more effective correction to the background based on the observations. If the ensemble fails to capture the space of uncertainties completely, the analysis cannot fully benefit from the observations regardless of their quality. 

The LETKF also has parameters that must be determined by ``tuning'' (experimentation). One such parameter is the number of ensemble members, which we choose to be 40 for all experiments. Another is the localization radius, which determines the distance within which observations are considered for the estimation of the state at a grid point. We use a localization radius of 1000\,km in the horizontal direction and $\sigma = 0.1$ in the vertical direction. Finally, all ensemble-based DA schemes must use some form of covariance inflation to compensate for the inevitable underestimation of the uncertainty in the knowledge of the state. The sources of this underestimation are forecast model errors, sampling errors due to the low number of ensemble members relative to the dimensionality of the dynamics, and nonlinearity of the evolution of the uncertainties \citep[e.g.,][]{Szunyogh2014}. We use the simplest form of covariance inflation, which is a multiplicative inflation with a scalar factor $\eta>1$. Such a covariance inflation factor can also be interpreted as a coupling parameter necessary for the synchronization of  the dynamics described by the analyses and the actual dynamics of the atmosphere \citep{Baek2004}; for an insufficiently high value of $\eta$, there can be occasional large bursts in the magnitude of the errors in a long time series of analyses. We tested values of $\eta$ from 1.2 to 2.1 for each experiment, but we present results only for the one value for which the global magnitude of the analysis error was found to be the lowest for the specific experiment. We provide the specific value of $\eta$ in the description of each experiment.

\subsection{The analysis-forecast experiments}
As already mentioned, the training period is from 0000\,UTC January 1, 1981 to 0000\,UTC January 2010 (the last training ``time step'' started at 1800\,UTC December 31, 2009). The testing period is from 0000\,UTC January 1, 2011 to 1800\,UTC December 31, 2011. We leave a one year gap between the end of the training period and the beginning of testing period to avoid seasonal and shorter term correlations between the ERA5 reanalyses used for training and testing. An analysis is prepared every 6 hours (at 0000\,UTC, 0600\,UTC, 1200\,UTC, and 1800\,UTC) and a 10-day forecast is started from each 0000\,UTC and 1200\,UTC analysis. We carry out the following four experiments:

\paragraph{PHYS experiment} No model training is done, because the analyses and forecasts of the testing period are prepared by using (the physics-based) SPEEDY as the forecast model. Results are presented for covariance inflation factor $\eta=1.9$. The purpose of this experiment is to provide benchmark verification statistics against which the improvements from hybridization can be assessed.

\paragraph{HYBRID-OPT experiment} The hybrid model  is trained on the ERA5 reanalyses for the training period. The analyses and forecasts of the testing period are prepared with this trained hybrid model as the forecast model. Results are presented for covariance inflation factor $\eta=1.9$. The purpose of this experiment is twofold. First, it simulates a hypothetical operational implementation of the hybrid model  in which a readily available high-quality reanalysis data set is used for model training and the offline trained model is used for real-time data assimilation and forecasting. Second, because the ERA5 reanalyses represent the known ``true'' atmospheric states in our experiments, this experiment also represents the (operationally unattainable) ideal situation in which the hybrid model is trained on a true state space trajectory of the atmosphere. Hence, the verification statistics from this experiment can serve as benchmarks to assess the effectiveness of the hybridization when imperfect analyses of the states are used for training, as in the next two experiments described below.

\paragraph{HYBRID-1 experiment} The hybrid model is trained on analyses prepared for the training period by SPEEDY as the forecast model of the DA process. The analyses and forecasts of the testing period are prepared with this trained hybrid model as the forecast model. 
Results are presented for covariance inflation factor $\eta=1.9$. The purpose of this experiment is to asses the analysis and forecast improvements that can be achieved by the hybridization of the forecast model when the training is done on analyses obtained by the physics-based model. Hypothetically, such a strategy could be followed by an operational center that has already prepared a reanalysis data sets with their physics-based model, in the hope that the hybridization of the same model would lead to improvements of their real-time analyses and forecasts. The testing period analyses obtained in this experiment, like the analyses in an online-training scheme, are prepared by a forecast model enhanced by hybridization.

\paragraph{HYBRID-2 experiment} The hybrid model is retrained on analyses prepared for the training period with the hybrid model trained on analyses with the physics-based model. The analyses and forecasts of the testing period are prepared with the retrained hybrid model as the forecast model. Results are presented for covariance inflation factor $\eta=1.7$. The design of this experiment is primarily motivated by the results of \citep{Wikner2021}, which were obtained for the Kuramoto-Sivashinsky system, a prototype spatiotemporally chaotic system. They found that retraining the hybrid model, iterating the DA for the training period and the model training that followed it, led to further modest analysis improvements. The training approach of this experiment is more similar to online training than that of the HYBRID-1 experiment since the analyses used for training can potentially also benefit from the hybridization of the forecast model.

\subsection{Verification statistics}
The following verification statistics are computed for each experiment for the one year long testing period:

\paragraph{Root-mean square error}
Let $\boldsymbol{z}(\sigma,t)$ be composed of the components of the global state vector $\boldsymbol{v}(t)$ for a single state variable (e.g., temperature) at vertical level $\sigma$ and time $t$. In addition, let superscripts $a$ indicate the analyses of one of the experiments and superscript $E$ denote the ERA5 reanalyses used for the verification of these analyses. We define the (global) root-mean-square error of the analysis $\boldsymbol{z^a}(\sigma,t)$ by
\begin{equation}
\epsilon_a \left( \boldsymbol{z}^a,\sigma,t \right) = \left( \frac{1}{4608} \sum_{i=1}^{4608} w_i \left[z_i^a(\sigma,t)-z_i^E(\sigma,t) \right]^2 \right)^{1/2}.
\end{equation}
In this equation, the index $i=1,2,\dots,4608(=96\times48)$ identifies the different components (horizontal grid point values) of $\boldsymbol{z}(t)$ and $w_i = \cos{\varphi_i}/\sum_{j=1}^{48} \varphi_j$ is a weight proportional to the area represented by the grid-point variable $z_i$, where $\varphi_i$ is the geographical latitude associated with $z_i$ and $\varphi_j$, $j=1,2,\dots,48$, are the different geographical latitudes of the model grid.

\paragraph{Mean vertical profile of the root-mean-square error}
We define the mean vertical profile of the root-mean-square error  by
\begin{equation}
\epsilon_\sigma \left( \boldsymbol{z}^a,\sigma \right) = \frac{1}{K} \sum_{k=1}^K  \epsilon_a \left( \boldsymbol{z}^a,\sigma,t_k \right), 
\end{equation}
where the mean is calculated over the $K=1460=(365\times4)$ verification times $t_k$ of the testing period.

\paragraph{Error maps}
Error maps are prepared by using the definition
\begin{equation}
\epsilon_{i}(\boldsymbol{z}^a,\sigma) = \left( \frac{1}{K} \sum_{k=1}^K \left[z_i^a(\sigma,t_k)-z_i^E(\sigma,t_k) \right]^2 \right)^{1/2}, 
\end{equation} 
$i=1,2,\dots,4608$, of the grid-point values of the root-mean-square error. To gain insights into the effects of hybridization on the systematic versus transient components of the errors, we also decompose the grid-point values $\epsilon^2_i (\boldsymbol{z}^a,\sigma)$ of the mean-square error as 
\begin{equation}
\epsilon^2_i (\boldsymbol{z}^a,\sigma) = B_i^2 (\boldsymbol{z}^a,\sigma) + \Sigma_i^2 (\boldsymbol{z}^a,\sigma), 
\end{equation}
$i=1,2,\dots,4608$, where
\begin{equation}
B_i (\boldsymbol{z}^a,\sigma) = \frac{1}{K} \sum_{k=1}^K \left[z_i^a(\sigma,t_k)-z_i^E(\sigma,t_k) \right],
\end{equation}
$i=1,2,\dots,4608$, is the systematic error (bias) and
\begin{equation}
\Sigma_i^2 (\boldsymbol{z}^a,\sigma) = \frac{1}{K} \sum_{k=1}^K \left[z_i^a(\sigma,t_k)-z_i^E(\sigma,t_k) - B_i (\boldsymbol{z}^a,\sigma) \right]^2 , 
\end{equation}
$i=1,2,\dots,4608$, is the error variance.

\paragraph{Forecast error growth curves}
Let $\boldsymbol{z}^f(\sigma,t_f,t)$ be the $t_f$ lead time global forecast of a single state variable at vertical level $\sigma$ for verification time $t$. We define the forecast error growth curve $\boldsymbol{z}^f(\sigma,t_f)$, $0 \le t_f \le 10$\,days, by
\begin{flalign}
& \epsilon_f \left( \boldsymbol{z}^f,\sigma,t_f \right) = & \\
& \frac{2}{\sqrt{4608}K} \sum_{k=1}^{K/2} \left( \sum_{i=1}^{4608} w_i \left[z_i^f(\sigma,t_f,t_k)-z_i^E(\sigma,t_f,t_k) \right]^2 \right)^{1/2}, & \nonumber
\end{flalign}
where the sample mean is calculated over the $K/2=730(=365\times2)$ forecast verification times $t_k$ of the testing period.

\section{Results of the experiments} \label{sec:results}

\subsection{Analysis performance of the hybrid model}

\subsubsection{Time series of the analysis error}
\begin{figure}[h]
\centerline{\includegraphics[width=0.5\textwidth]{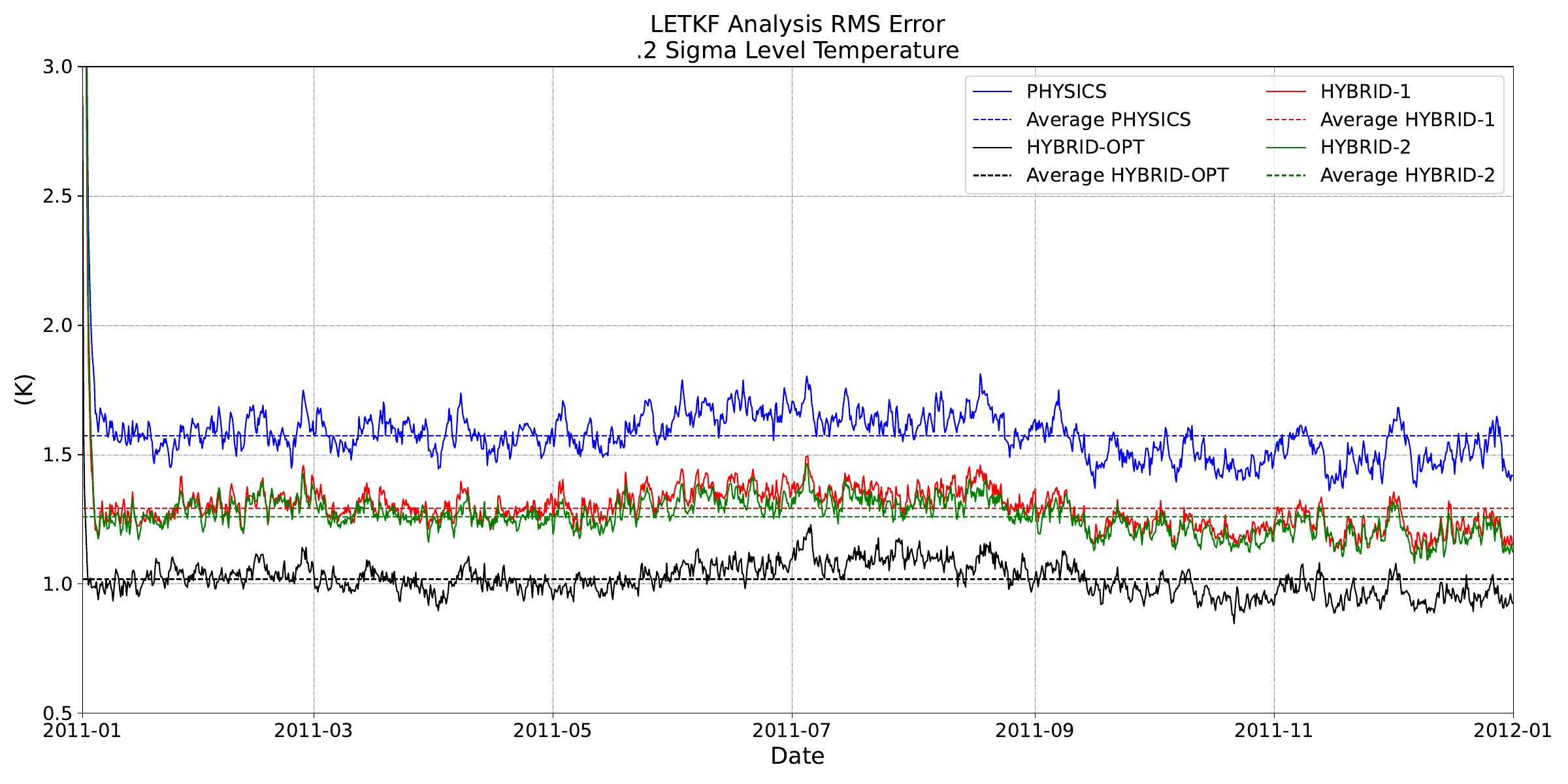}}
\caption{Temporal evolution of the root-mean-square analysis error for the temperature at $\sigma=0.2$.}
\label{fig:2}
\end{figure}

\begin{figure}[h]
\centerline{\includegraphics[width=0.5\textwidth]{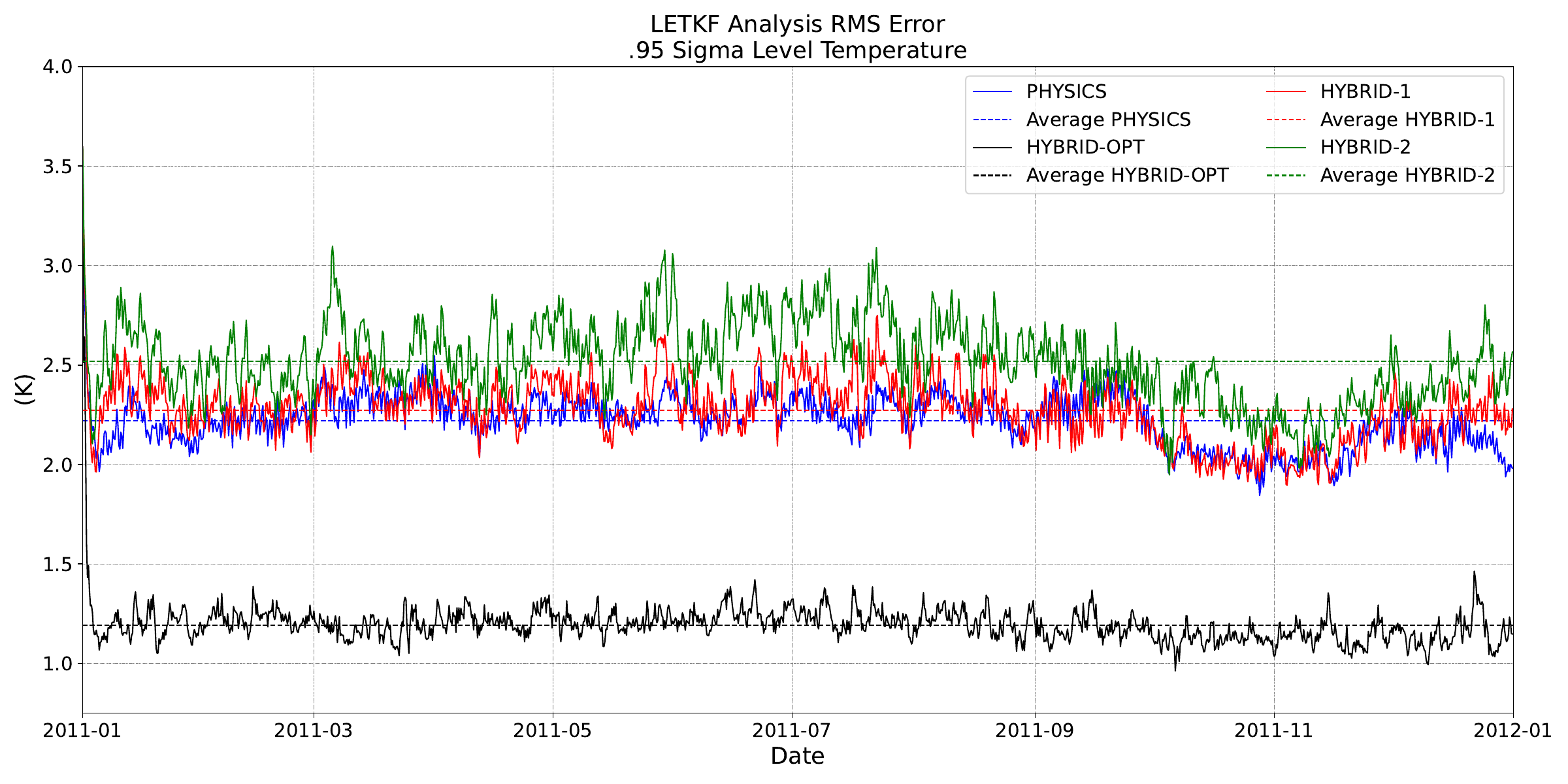}}
\caption{Same as Fig.~\ref{fig:2}, except at $\sigma=0.95$.}
\label{fig:3}
\end{figure}

After a brief transient period of about 6 days, the values of the globally averaged analysis errors $\epsilon_a \left( \boldsymbol{z}^a,\sigma,t \right)$ vary around a stable mean
for all state variables and $\sigma$ levels. This behavior is illustrated with the results for the temperature at $\sigma=0.2$ (Fig.~\ref{fig:2}) and  $\sigma=0.95$ (Figure~\ref{fig:3}). In these figures, each dashed line indicates the (temporal) mean of the values shown by the solid line of the same color  for one of the experiments (blue: PHYS, black: HYBRID-OPT, red: HYBRID 1, green: HYBRID-2). The percent values of the error reduction shown in Table~\ref{tab:1} are based on the values $\epsilon_a \left( \boldsymbol{z}^a,\sigma\right)$ associated with these dashed lines. At $\sigma=0.2$, all three configurations of the hybrid model lead to substantial reduction of the mean analysis error: 35.3\% for HYBRID-OPT, 17.9\% for HYBRID-1 and 19.9\% for HYBRID-2. This is the type of behavior we have been hoping for: the hybridization leads to a substantial reduction of the magnitude of the analysis errors when the model is trained on ERA5 reanalyses, more than half of that reduction is retained when the model is trained on imperfect analyses prepared with the physics-based model, and retraining the model leads to a modest further reduction of the magnitude of the errors. The situation, however, is very different at the lowest model level, where the hybridization of the model leads to an even larger 46.3\% reduction of the magnitude of the analysis errors when the training is done on ERA5 reanalyses, but this error reduction turns into a 2.4\% increase of the error when the training is done on analyses obtained with the physics-based model, and retraining the model on the analyses of the HYBRID-1 analyses results in an even larger, 13.5\%, increase of the error.  
\begin{table}[h]
\caption{The reduction of the temporal mean of the root-mean-square of the analysis errors for the prognostic state variables at vertical levels $\sigma=0.2$, $\sigma=0.51$, and $\sigma=0.95$. Positive values indicate reduction while negative values indicate degradation of the errors compared to those for the PHYS experiment.}\label{tab:1}
\begin{center}
\begin{tabular*}{0.48\textwidth}{ccccrrcrc}
\topline
{\scriptsize State variable} & {\scriptsize HYBRID-OPT} & {\scriptsize HYBRID-1} & {\scriptsize HYBRID-2}\\
\midline
 T($\sigma=0.2$) & 35.3\% & 17.9\% & 19.9\% \\
 q($\sigma=0.2$) & 75.4\% & 2.2\% & -0.4\% \\
 v($\sigma=0.2$) & 21.8\% & 12.0\% & 13\% \\
 u($\sigma=0.2$) & 22.5\% & 10.8\% & 11.9\% \\
 T($\sigma=0.51$) & 22.7\% & 2.8\% & 1.5\% \\
 q($\sigma=0.51$) & 15\% & 4.2\% & 7.3\% \\
 v($\sigma=0.51$) & 14.3\% & 6.7\% & 6.4\% \\
 u($\sigma=0.51$) & 14.8\% & 6.1\% & 5.9\% \\
 T($\sigma=0.95$) & 46.3\% & -2.4\% & -13.5\% \\
 q($\sigma=0.95$) & 48.5\% & 0.1\% & -8.3\% \\
 v($\sigma=0.95$) & 32.9\% & 1.4\% & -1.1\% \\
 u($\sigma=0.95$) & 34.7\% & 1.3\% & -1.4\% \\
 $p_s$ & 93\% & -0.1\% & -0.5\% \\
\botline
\end{tabular*}
\end{center}
\end{table}

The general trend that emerges from Table~\ref{tab:1}  is that the hybridization leads to a very substantial improvement of the analysis accuracy when the model is trained on the true past trajectory of the atmosphere, but the improvements can be modest, or may even turn into a degradations when the model is trained on analyses obtained by using the physics-based model to provide the backgrounds. Retraining the model leads to the anticipated further improvements of the analysis accuracy only in cases in which the original training has already led to a substantial improvement. The atypical values for the surface pressure reflect the fact that the model forecasts that provide the background ensemble for the LETKF in the PHYS, HYBRID-1, and HYBRID-2 experiment have large systematic errors because of the low-resolution orography of the model. In fact, the global surface pressure error varies very little around a mean of 17.5\,hPa in these experiments, which suggests that the error is dominated by the effect of the orography difference between the model and the verification data set. In the HYBRID-OPT experiment, the corresponding mean is 1.3~hPa, which leads to the 93\% error reduction shown in the table. 


\subsubsection{Mean vertical profiles of the errors}
\begin{figure}[h]
\centerline{\includegraphics[width=0.5\textwidth]{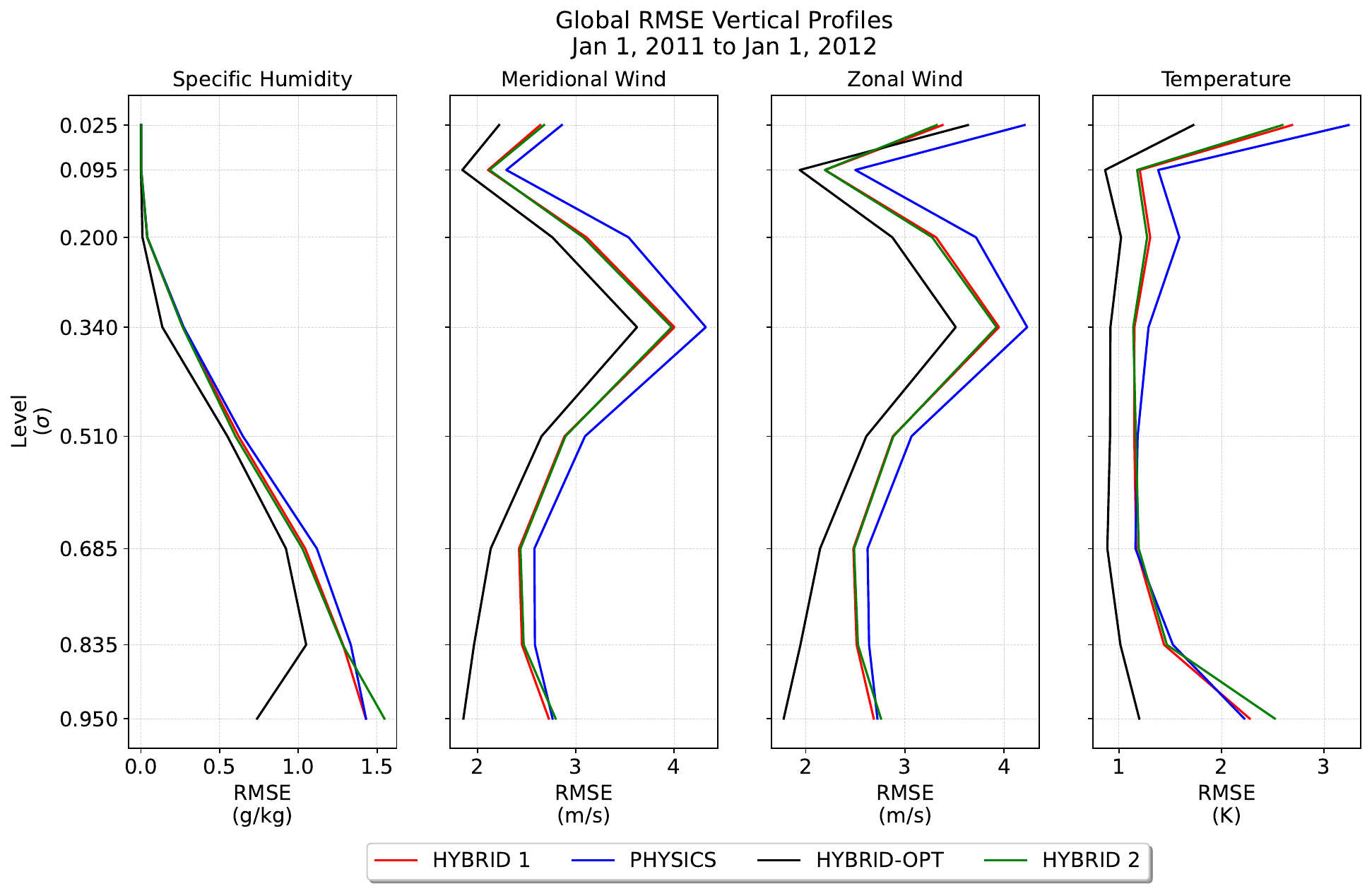}}
\caption{Vertical profiles of analysis errors for the four experiments state. Shown are the values of $\epsilon_\sigma \left( \boldsymbol{z}^a,\sigma \right)$ (from left to right) for the specific humidity, meridional wind component, zonal wind component, and temperature.}
\label{fig:4}
\end{figure}

The results shown thus far suggest that the benefits of the hybridization of the forecast model for the analyses strongly depend on the vertical level. For the further investigation of this behavior, Fig.~\ref{fig:4} shows the $\epsilon_\sigma \left( \boldsymbol{z}^f,\sigma \right)$ vertical profiles of the root-mean-square error for the different state variables. The results shown in the figure confirm that the hybridization of the forecast model leads to substantial reduction of the analysis errors for all state variables and vertical levels when the training is done on ERA5 reanalyses. The only exception is the specific humidity at the top two model levels, where the DA with the physics-based model also correctly captures that its value is nearly zero. The figure also demonstrates that the negative results shown earlier for $\sigma=0.95$ are the exceptions, because there are no degradations from the hybridization at any other level in the HYBRID-1 or HYBRID-2 experiment. In addition, there are clear improvements throughout the entire atmospheric column for the two wind components, the specific humidity below $\sigma=0.51$, and the temperature above $\sigma=0.51$. The negligible differences between the profiles for the HYBRID-1 or HYBRID-2 experiment suggest, however, that there are no obvious benefits of retraining the model on the analyses of the HYBRID-1 experiment.

\subsubsection{Error maps}
\begin{figure}[h]
\centerline{\includegraphics[width=0.5\textwidth]{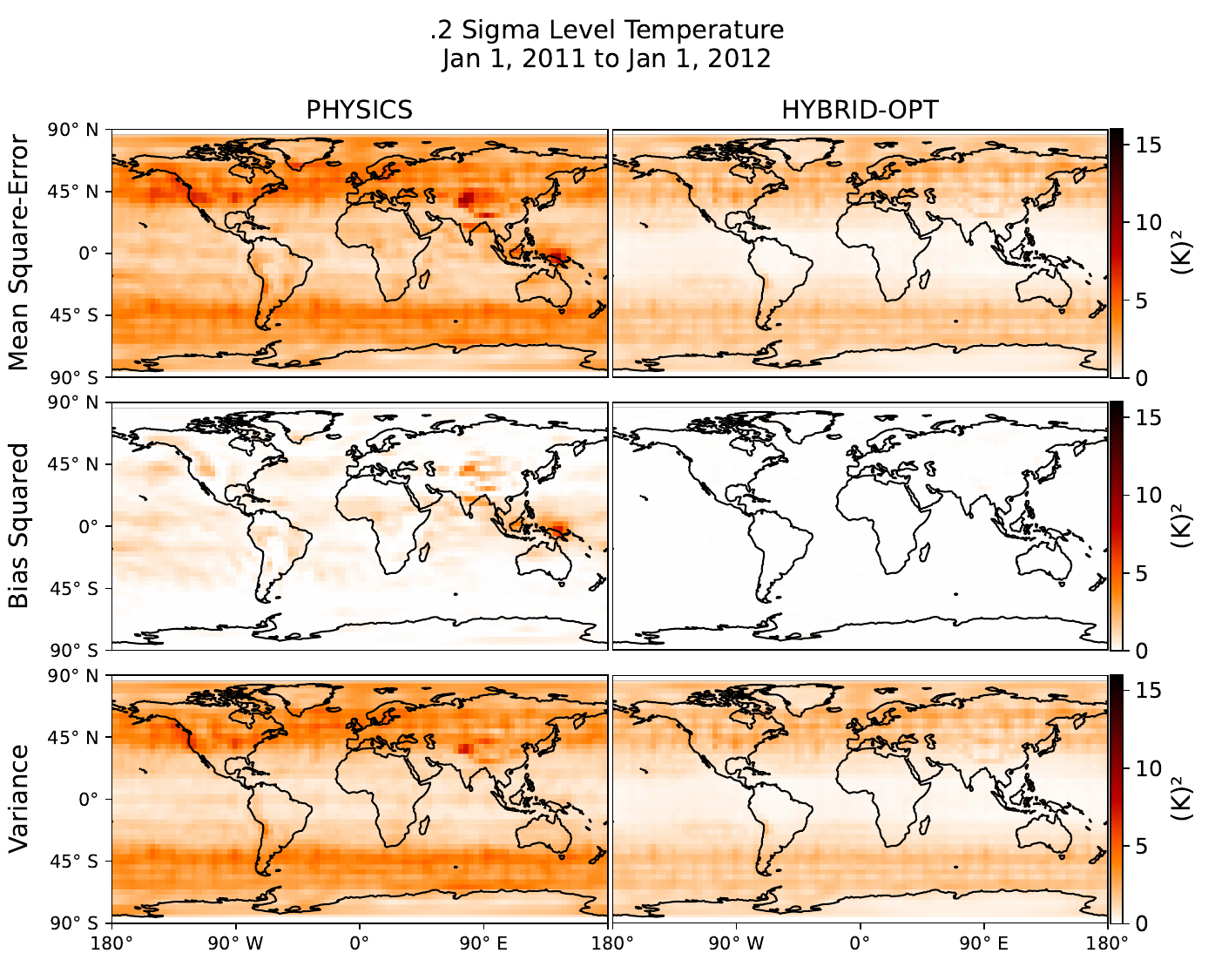}}
\caption{Maps of the mean-square analysis error and its decomposition for the temperature at vertical level  $\sigma=0.2$. Shown are (top) the mean square error, (middle) the square of the bias, and (bottom) the error variance for the (left) PHYS and (right) HYBRID-OPT experiment.}
\label{fig:5}
\end{figure}

\begin{figure}[h]
\centerline{\includegraphics[width=0.5\textwidth]{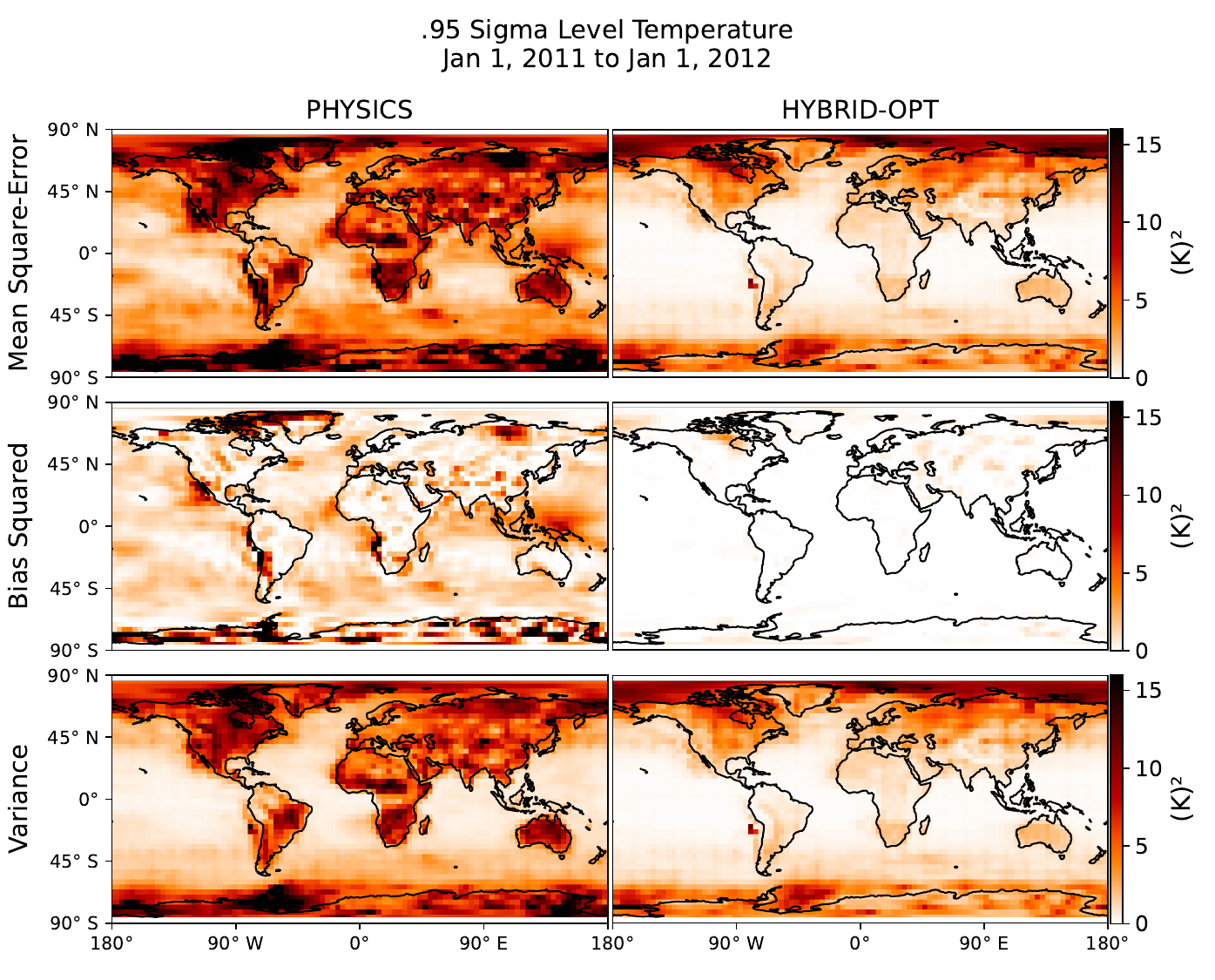}}
\caption{Same as Fig.~\ref{fig:5}, except for the temperature at vertical level  $\sigma=0.95$.}
\label{fig:6}
\end{figure}

\begin{figure}[h]
\centerline{\includegraphics[width=0.5\textwidth]{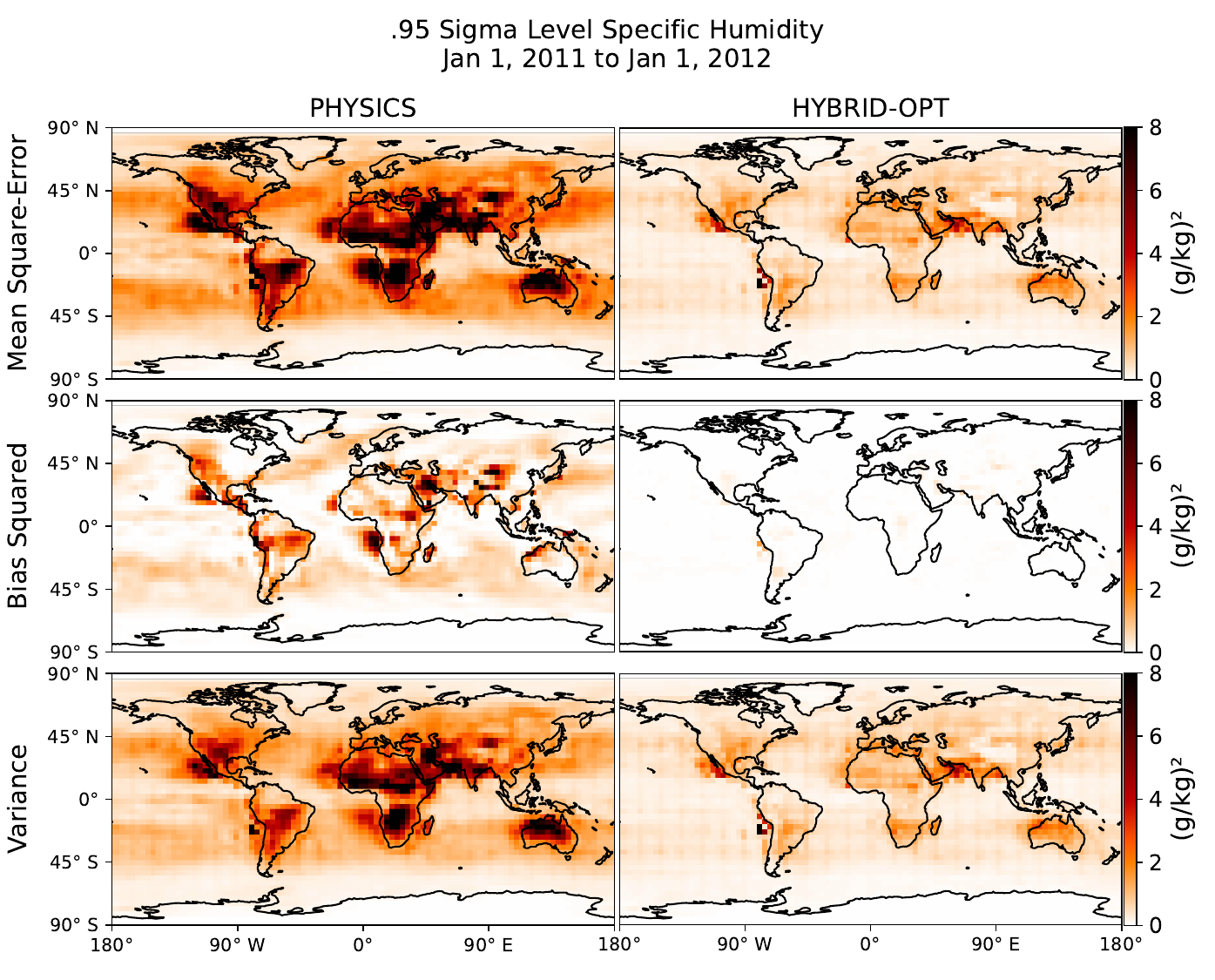}}
\caption{Same as Fig.~\ref{fig:5}, except for the specific humidity at vertical level  $\sigma=0.95$.}
\label{fig:7}
\end{figure}

\begin{figure}
\centerline{\includegraphics[width=0.5\textwidth]{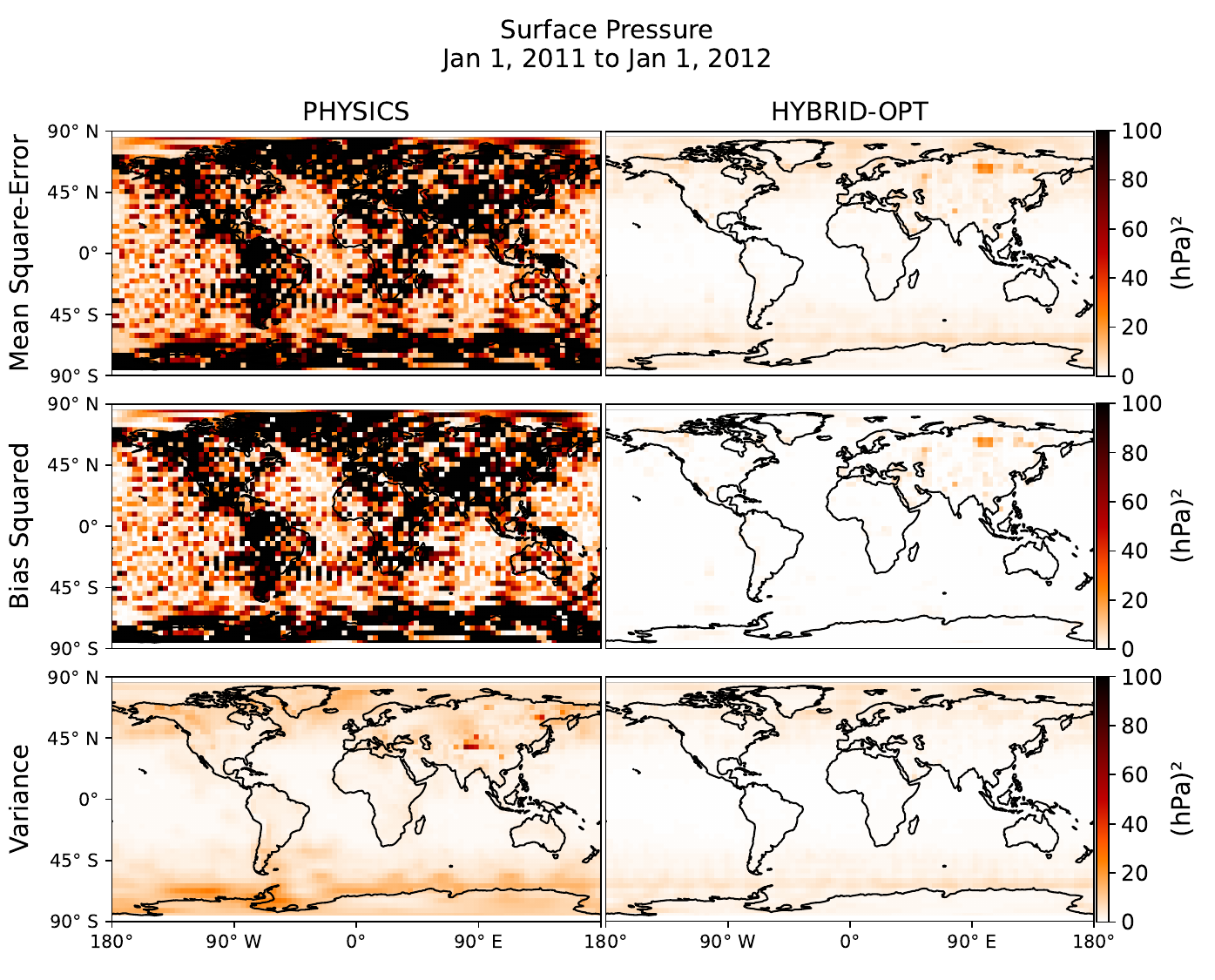}}
\caption{Same as Fig.~\ref{fig:5}, except for the surface pressure.}
\label{fig:8}
\end{figure}

We start the discussion of the error maps with a comparison of the results of the PHYS and HYBRID-OPT experiment. Figures \ref{fig:5}, \ref{fig:6}, \ref{fig:7}, and \ref{fig:8} illustrate the most important general trends in the results for this pair of experiments: the hybridization almost completely eliminates the systematic errors (bias) and greatly reduces the magnitude of the transient errors (error variance), leading to a substantial reduction of the root-mean-square error. The specific examples shown in the four figures were selected to illustrate the different ways hybridization of the forecast model can improve the analysis performance. Specifically, (Fig.~\ref{fig:5}) shows that the hybridization 
improves the analysis of the temperature at $\sigma=0.2$, because the hybrid model has a realistic atmospheric dynamics at the top model levels in contrast to SPEEDY. For instance, \citet{Arcomano2022} showed that the hybridization greatly reduced the magnitude of the temperature bias at the 200\,hPa level (from a maximum of about 9\,K to a maximum of about 2\,K), while \citet{Arcomano2023} demonstrated that the model was able to produce realistic sudden stratospheric warming events at the 25\,hPa pressure level. The reduction of the model bias helps the LETKF, which in the configuration used in the present study assumes no background bias, correctly interpret the observations. Thus the elimination of the background bias sets the stage for the reduction of the analysis error variance. This reduction could not be materialized, however, if the background ensemble would not be able to capture at least some important features of the uncertainty dynamics. (Recall that the LETKF can make corrections of the estimate of the state based on the observations only in the space spanned by the background perturbations.) Hence, the reduction of the analysis error variance indicates that the hybridization improves the performance of the model in capturing the dynamics of the forecast (background) uncertainties . 

At the lowest model level ($\sigma=0.95$), the simplified parameterization schemes of SPEEDY are the main sources of the model errors. These errors lead to large local values of the bias and error variance in the analyses of the temperature (Fig.~\ref{fig:6}) and the specific humidity (Fig.~\ref{fig:7}) in the PHYS experiment. The hybridization of the forecast model almost completely eliminates these biases and greatly reduces the magnitude of the transient errors over land at the low- and mid-latitudes (e.g., South America, Africa, Australia). 

As mentioned earlier, the surface pressure is a special variable because of the inevitable large local biases introduced in the mountainous regions in SPEEDY. Figure~\ref{fig:8} shows that this bias dominates the local values of the surface pressure analysis error. It also shows that the hybridization is highly effective in reducing (nearly eliminating) this bias, enabling the hybrid model to make a substantial reduction of the analysis variance as well.

\begin{figure}[h]
\centerline{\includegraphics[width=0.5\textwidth]{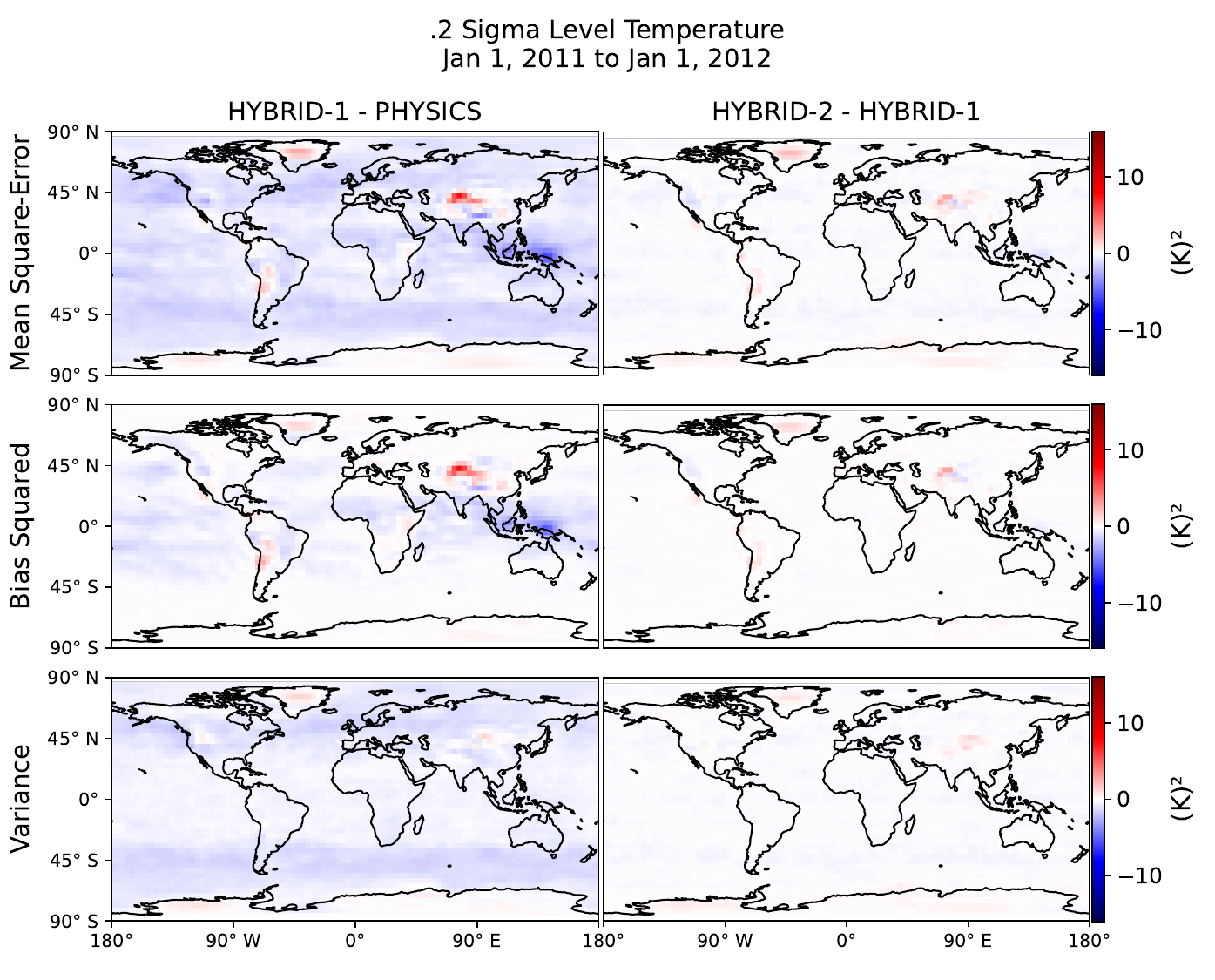}}
\caption{Maps of the differences in the different analysis error components between pairs of experiments for the temperature at vertical level  $\sigma=0.2$. Shown are the differences in (top) the mean square error, (middle) the square of the bias, and (bottom) the error variance between the (left) HYBRID1 and PHYS experiment, and (right) HYBRID-2 and HYBRID-1 experiment.}
\label{fig:9}
\end{figure}

\begin{figure}[h]
\centerline{\includegraphics[width=0.5\textwidth]{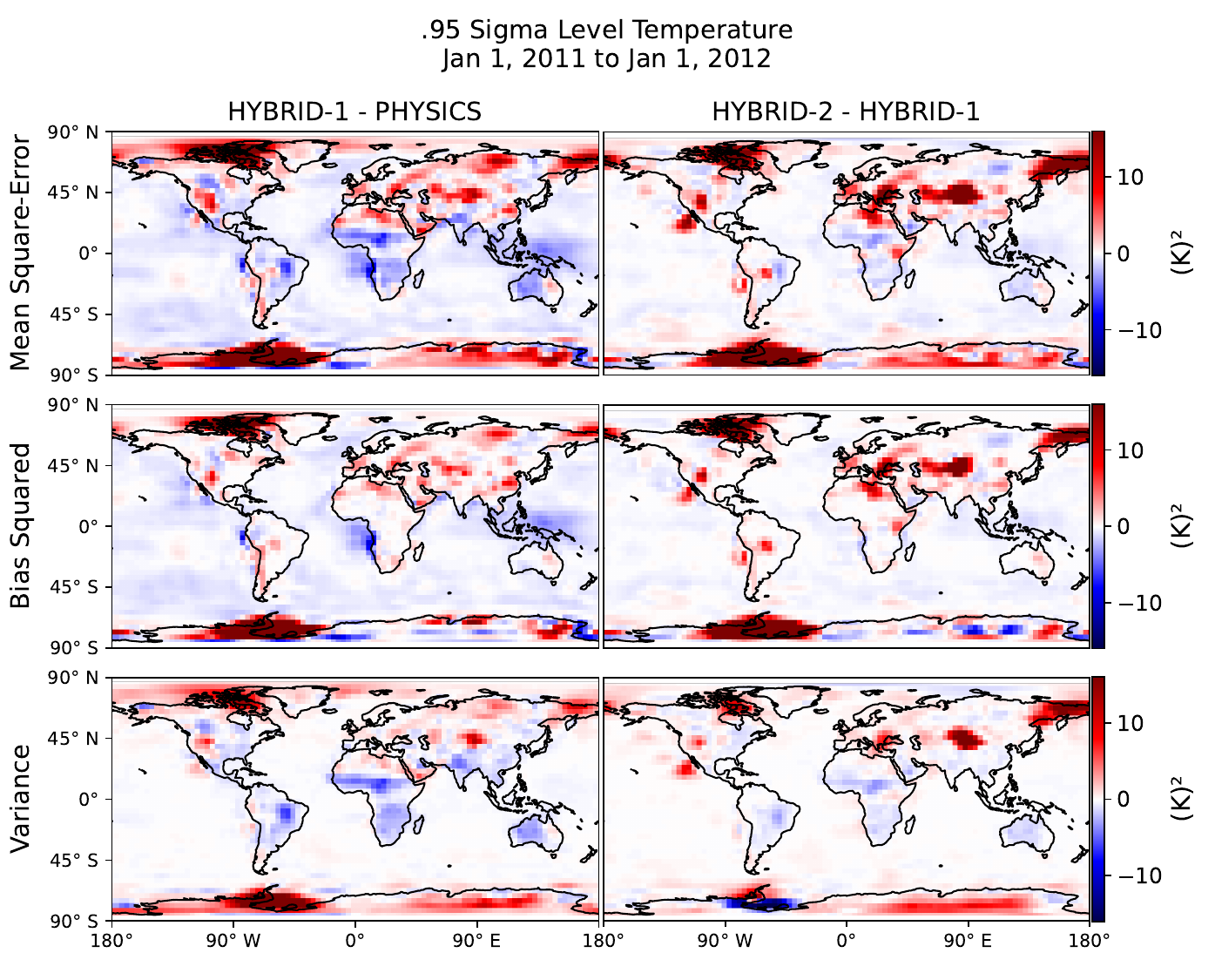}}
\caption{Same as Fig.~\ref{fig:9}, except for the temperature at vertical level  $\sigma=0.95$.}
\label{fig:10}
\end{figure}

\begin{figure}[h]
\centerline{\includegraphics[width=0.5\textwidth]{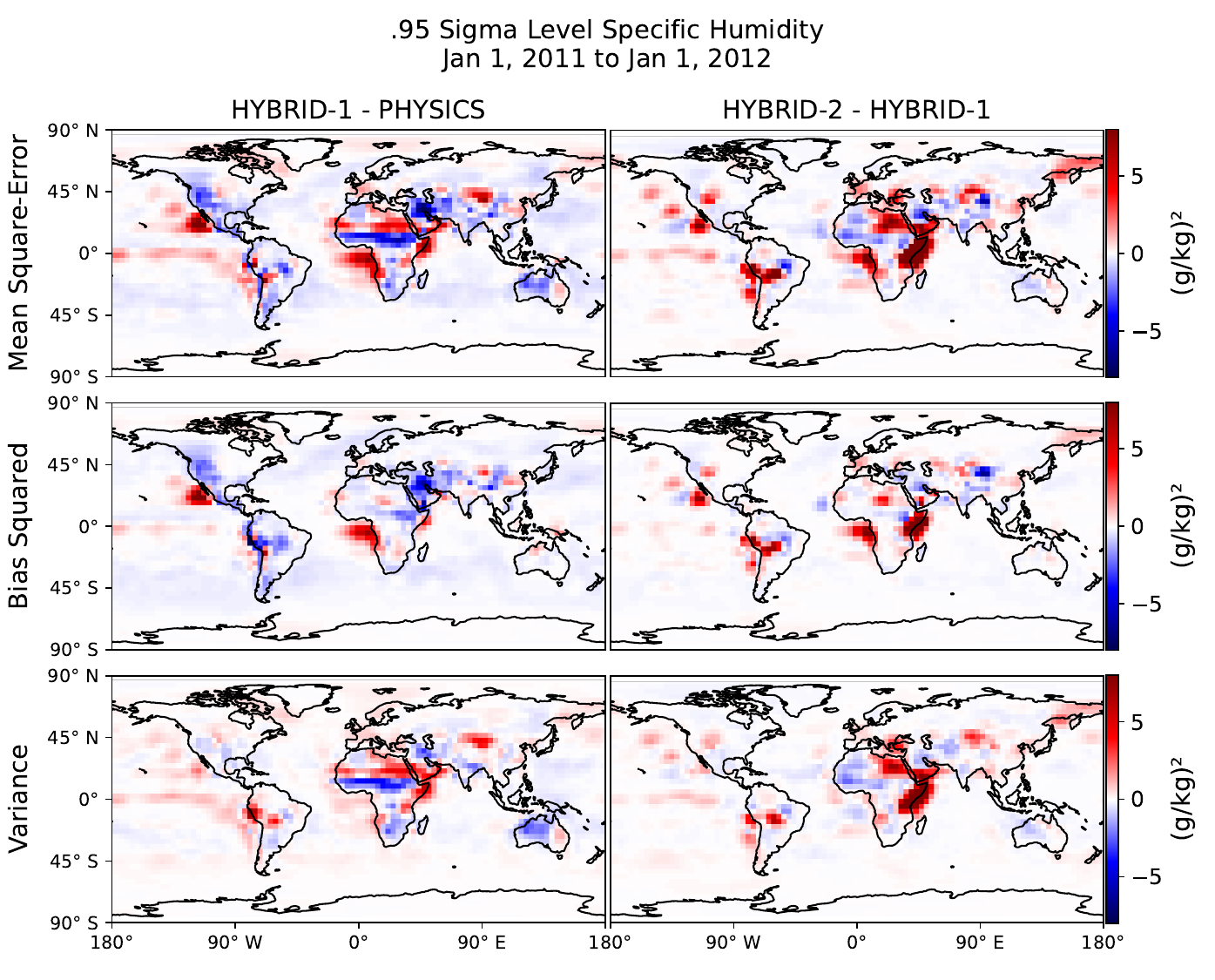}}
\caption{Same as Fig.~\ref{fig:9}, except for the specific humidity at vertical level  $\sigma=0.95$.}
\label{fig:11}
\end{figure}

For the comparison of the spatial patterns of the errors in the PHYS, HYBRID-1, and HYBRID-2 experiments, a different format is used in Figs.~\ref{fig:9},  \ref{fig:10}, and \ref{fig:11} to show the error maps than in the earlier figures: these figures show the differences between the errors rather than the errors themselves for pairs of the experiments. The differences are shown for the HYBRID-1 and PHYS experiment (left panels), and the HYBRID-2 and HYBRID-1 experiment (right panels). Blue shades indicate that the HYBRID-1 analyses are more accurate than the PHYS analyses (left panels) and the HYBRID-2 analyses are more accurate than the HYBRID-1 analyses (right panels). Red shades indicate the opposite outcomes. Ideally, we would see only blue shades in these figures, but the figures show more mixed results. For the temperature at $\sigma=0.2$ (Fig.~\ref{fig:9}), the results are almost uniformly positive for the HYBRID-1 experiment versus the PHYS experiment (left panels), which suggests that the hybridization helps the analyses in the upper troposphere even if the model is trained on analyses obtained with the physics-based model. The few exceptions are the small local increases in the bias in regions of high orography (Himalayas, Andes, Greenland), which affect the analysis variance only slightly. Interestingly, the DA with the hybrid model is even able to reduce the relatively large local bias over Indonesia (Fig.~\ref{fig:5}). Retraining the model on analyses obtained with the hybrid model has little effect on the analysis accuracy (right panels of Fig.~\ref{fig:9}), but in the regions where the magnitude of the bias is larger for the HYBRID-1 experiment than for the PHYS experiment, the retraining tend to lead to a small further increase of the bias.

The results are more mixed at  $\sigma=0.95$ (Figs.~\ref{fig:10} and \ref{fig:11}), which is not unexpected based on the overall results described for that level earlier (Fig.~\ref{fig:3}, Fig.~\ref{fig:4}, and Table~\ref{tab:1}). The two figures show that the small changes in the overall accuracy are the results of offsetting localized improvements and degradations that can have considerable magnitudes. The large local degradations of the accuracy of the temperature analyses (Fig.~\ref{fig:10}) are likely to be the result of the crude handling of the pole problem in the hybrid model and the difficulties of the RC component of the model to make corrections to the physics-based forecasts over ice surfaces. (Fig.~\ref{fig:6} shows that the hybrid model has difficulties in these regions even if the model is trained on ERA5 reanalyses.) These errors are only amplified when the hybrid model is retrained on analyses of the HYBRID-1 experiment (right panels of Fig.~\ref{fig:10}), and the retraining has a particularly negative effect on the biases (middle right panel of Fig.~\ref{fig:10}). The main regions of improvements are over the continents in the SH (with the exception of Antartica), where the variance of the analysis errors is reduced in the HYBRID-1 experiment (bottom left panel of Fig.~\ref{fig:10}) and further reduced in the HYBRID-2 experiment. 

Maps are not shown for the surface pressure for the HYBRID-1 and HYBRID-2 experiment, because the hybridization has little effect on these maps compared to those shown for the PHYS experiments (left panels of Fig.~\ref{fig:8}). This result shows that the analyses prepared with the hybrid model trained on analyses with the physics-based model are just as vulnerable to the large surface pressure biases associated with the model orography as the analyses prepared with the physics-based model. Retraining cannot fix this problem, which is not surprising considering the fact that retraining does not help reduce the biases for the other variables either. One potential solution to address this problem would be to do an online correction of the surface pressure bias in the observation operator of the DA. Such a bias correction was found to be highly effective for SPEEDY in \cite{Baek2009}. To use this approach, the current configuration of the hybrid model would not have to be changed. Another option would be to add the ERA5 orography to the input parameters of the hybrid model. The model would hopefully learn to use this information to make an effective correction of the surface pressure bias. While this approach has not been tested, yet, if it worked, it could effectively reduce the surface pressure analysis bias without the modification of the DA code. This could, in turn, reduce the analysis errors for other variables near the surface.

\subsection{Forecast performance of the hybrid model}
\begin{figure}[h]
\centerline{\includegraphics[width=0.5\textwidth]{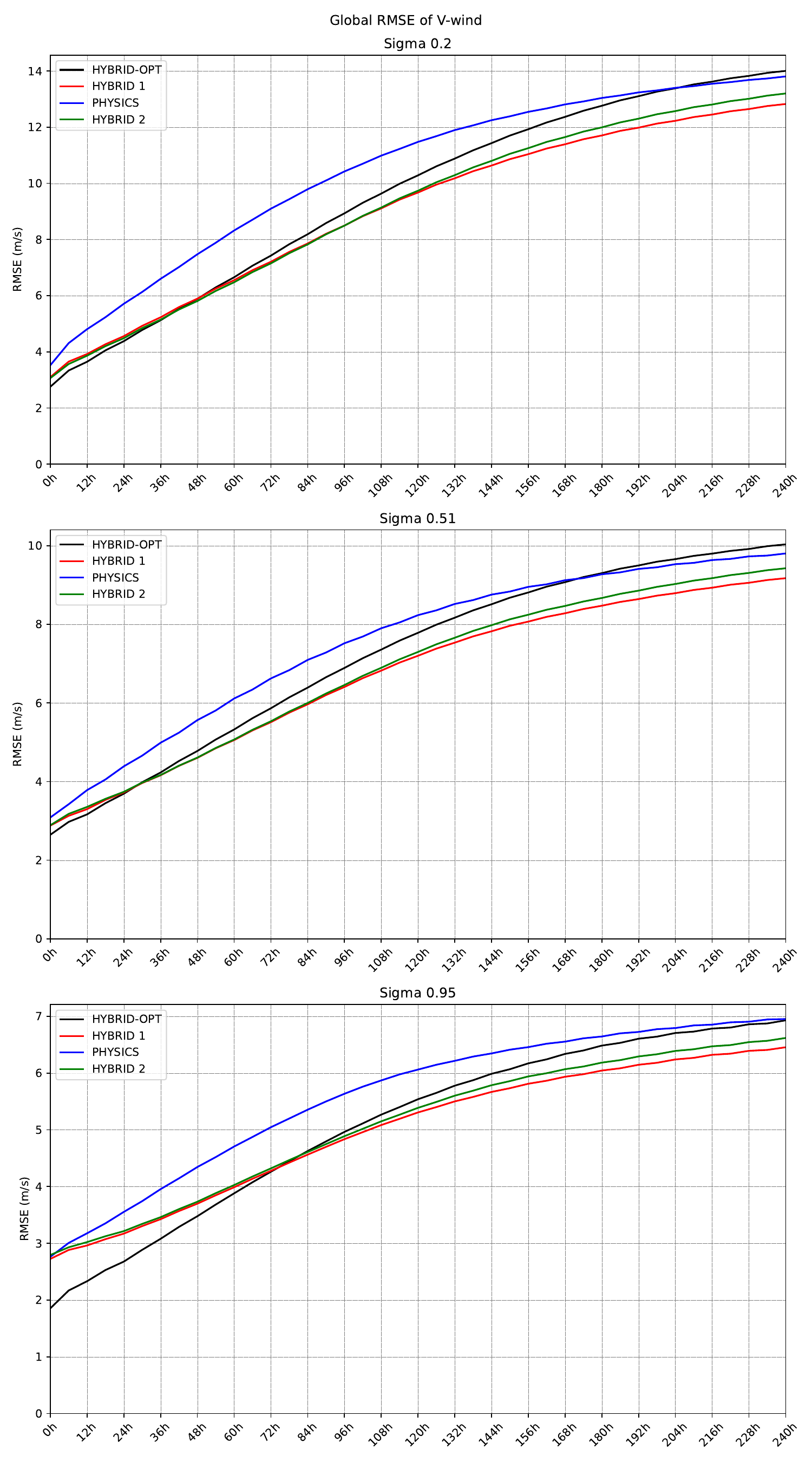}}
\caption{The mean-square error of the meridional wind forecasts as function of the forecast lead time at three selected sigma levels. Shown are the error growth curves for (top) $\sigma=0.2$, (middle) $\sigma=0.51$, and (bottom) $\sigma=0.95$.}
\label{fig:12}
\end{figure}

The behavior of the forecast error growth curves, $\boldsymbol{z}^f(\sigma,t_f)$, $0 \le t_f \le 10$\,days, is illustrated by the results for the meridional component of the wind (Fig.~\ref{fig:12}). 
The results show that the forecasts of the three hybrid model experiments are substantially more accurate than those of the PHYS experiment. For example, the forecasts of the HYBRID-OPT experiment at $\sigma=0.95$ are as accurate at 96\,h forecast time as the forecasts of the PHYS experiment at 70\,h forecast time. 
Interestingly, the forecasts of  the HYBRID-1 and HYBRID-2 experiment become more accurate than those of the HYBRID-OPT experiment after a few days (3~days at $\sigma=0.95$, 1~day at $\sigma=0.95$, and 2~days at $\sigma=0.95$).
The same two features are also present in the results (not shown) for the zonal wind, temperature, and specific humidity. The hybrid forecasts become more accurate than the forecasts of the PHYS experiment quickly (after 12-24~hours) even for the variables and vertical levels for which their analysis errors are 
Like the analysis errors, the forecast errors behave differently for the surface pressure than the other variables. Specifically, in the PHYS, HYBRID-1 and HYBRID-2 experiment, the error is slowly growing from 17.6~hPa at analysis time to about 19~hPa at 10~days forecast time. In contrast, in the HYBRID-OPT experiment, the error grows from 1.3~hPa to 6.2~hPa, which reflects the fact that surface pressure bias remains negligible at all forecast times. Interestingly, the results for the other variables suggest that the near elimination the orography-related surface bias does not have a sustained positive effect on the hybrid model forecasts.

To further investigate the forecast results in a more quantitative manner, we fit a hyperbolic tangent function
\begin{equation} \label{eq:Zagar}
\hat{\epsilon}(t_f)= A \tanh{ (at_f+b)} + B,
\end{equation} 
to the forecast error growth curves following \cite{zagar2017}. In Eq.~(\ref{eq:Zagar}), $A$, $B$, $a$, and $b$ are the real scalar parameters to be determined by function fitting. It can be shown that the Lorenz-curve $d \hat{\epsilon}/dt(\hat{\epsilon})$, which describes the dependence of the rate of the error growth on the magnitude of the error \citep{Lorenz1982} for (\ref{eq:Zagar}), satisfies the error growth model 
\begin{equation} \label{eq:RMS_DK}
\frac{d \hat{\epsilon}}{d t_f} = (\alpha \hat{\epsilon} + \beta) \left( 1 - \frac{\hat{\epsilon}}{\hat{\epsilon}_{max}} \right)
\end{equation}
of \cite{dalcher1987} with
\begin{equation} \label{eq:Zagar_DK}
\alpha=\frac{a}{A}(A+B), \ \ \beta= -\frac{a}{A} (A+B)(B-A), \ \ \hat{\epsilon}_{max}=(A+B).
\end{equation}
The standard interpretations of these parameters are the following: $\alpha$ is an estimate of the exponential growth rate of small values of $\epsilon$ 
by the chaotic dynamics, $\beta$ is an estimate of the rate of the contribution of model errors to the forecast error growth, and $\hat{\epsilon}_{max}$ is an estimate of the saturation value of $\epsilon$ that is reached when the forecasts completely lose their dependence on the state of the atmosphere at the beginning of the forecasts. This interpretation of $\alpha$ and $\beta$, however, is not without limitations. First, the forecast error growth is highly scale dependent: below the synoptic scales, smaller scale errors saturate earlier and at a lower level than the larger scale errors. For example, \cite{Arcomano2022} found that for the versions of the model used in the PHYS and HYBRID-OPT experiment of the present paper, the errors at global wave numbers higher than 20 saturate by the end of the first forecast day. As \cite{zagar2017} pointed out, these rapidly saturating smaller scale errors also contribute to $\beta$. Second, if the magnitude of the biases is comparable to the magnitude of the transient errors at initial and short forecast times, it can lead to even negative values of $\beta$. In such a situation, $\beta$ obviously cannot be interpreted as a measure of the forecast error. Finally, we note that Eq.~\ref{eq:RMS_DK} can also be written in the equivalent form 
\begin{equation} \label{eq:RMS_us}
\frac{d \hat{\epsilon}}{d t_f} = -c_2 \hat{\epsilon}^2 + c_1 \hat{\epsilon} +\beta,
\end{equation}
where $c_2=\alpha/\hat{\epsilon}_{max}$ and $c_1=\alpha - \beta/ \hat{\epsilon}_{max}$. The interpretation of $c_1$ and $c_2$ is less ambiguous than that of $\alpha$ and $\beta$ and we will refer to them as the effective linear growth rate and the rate of nonlinear saturation, respectively.

We fit the function of Eq.~(\ref{eq:Zagar}) to the 41 data points provided by the six-hourly values $\epsilon(t_f)$, $t_f=0,1,2,\dots,10$\,days of the root-mean-square error. The function fits the data points for all curves of Fig.~\ref{fig:12} accurately: both the $R^2$ values and the $R^2$ values adjusted for the difference between the number of fitted data points and parameters are equal to, or larger than, 0.999. In addition, the root-mean-square of $\epsilon (t_f) -\hat{\epsilon}(t_f)$ for the 11 data points is about 1-1.5\% of the smallest fitted value for each curve. We choose the curves for $\sigma=0.51$ for our analysis, because the biases are the smallest at this level in the four experiments. In contrast, at $\sigma=0.95$, the forecasts of the PHYS, HYBRID-1, and HYBRID-2 experiments have persistent biases in the mountainous regions (most prominently, in the Andes and Himalayas) that do not grow in magnitude are already present in the analyses (at initial forecast time). The fact that these biases make a substantial contribution to the root-mean-square errors at the early forecast times explains 
the slow error growth at those forecast times seen in Fig.~\ref{fig:12} for the HYBRID-1 and HYBRID-2 experiment. The same biases are also present in the PHYS experiment, but the more rapidly growing transient errors of that experiment make the effect of the biases on the overall growth rate less pronounced.  

The values of the estimated parameters and errors of the curve fitting for $\sigma=0.51$ are summarized in Table~\ref{tab:2}.
\begin{table}[h]
\caption{Estimated parameters of the forecast error growth curves for $\sigma=0.51$ shown in Fig.~\ref{fig:12}. The last column shows that values of root-mean-square of $\epsilon (t_f) -\hat{\epsilon}(t_f)$.} \label{tab:2}
\begin{center}
\begin{tabular*}{0.5\textwidth}{ccccccc}
\topline
 {\tiny Experiment} & {\tiny $\alpha$  [1/d]} & {\tiny $\beta$ [m/sd]} & {\tiny $\hat{\epsilon}_{max}$ [m/s]} & {\tiny $c_2$ [s/md]} & {\tiny $c_1$ [1/d]} & {\tiny Fit [m/s]} \\
\midline
{\tiny PHYS} & {\scriptsize 0.25} & 1.11 & 10.13 & 0.02 & 0.14 & 0.03 \\
{\tiny HYBRID-OPT} & 0.35 & 0.38 & 10.52 & 0.03 & 0.24 & 0.04 \\
{\tiny HYBRID 1} & 0.40 & -0.08 & 9.57 & 0.04 & 0.41 & 0.04 \\
{\tiny HYBRID 2} & 0.47 & -0.41 & 9.79 & 0.05 & 0.51 & 0.05 \\
\botline
\end{tabular*}
\end{center}
\end{table}
The value of $\alpha$ for sigma level 0.51 for the HYBRID-OPT experiment (0.35\,day$^{-1}$) is very similar to those that we obtained for a comparison with the physics-based ECMWF IFS (0.34\,day$^{-1}$) and ML-based ECMWF AIFS (0.34\,day$^{-1}$), GraphCast (0.34\,day$^{-1}$), Pangu-Weather (0.29\,day$^{-1}$), and FourCastNet (0.29\,day$^{-1}$) models based on the forecast error growth curves published for the 500~hPa geopotential height for December-January-February 2024-2025 at www.ecmwf.int.\footnote{Captured on June 3 2025.} The value of $\alpha$ (0.25\,day$^{-1}$) is somewhat lower for the PHYS experiment than the state-of-the-art forecast models. 

The growing advantage of the HYBRID-OPT forecasts over the PHYSICS forecasts in the first three forecast days is the result of the smaller value of $\beta$ (0.38\,m\,s$^{-1}$\,day$^{-1}$ vs. 1.11 \,m\,s$^{-1}$\,day$^{-1}$), which suggests that the hybridization of the model of the HYBRID-OPT experiment reduces the contribution of the model errors. The advantage of the HYBRID-OPT forecasts gradually decreases between forecast times 3 days and 7 days as a result of the higher saturation value $\hat{\epsilon}_{max}$ (10.52\,m\,s$^{-1}$ vs. 10.13\,m\,s$^{-1}$) of the root-mean-square error for the HYBRID-OPT forecasts. This result shows that the earlier advantage of these forecasts is not the result of a gradual smoothing (reduction of the spatial variance) of the meridional velocity field. This finding is not unexpected based on the results of \citet{Arcomano2022} that showed that the variance of the meridional wind field at the highest resolved wave numbers was higher for the ERA5-trained hybrid model than the physics-based SPEEDY. The reason for this behavior is that unlike a physics-based model, the ML component of the hybrid model does not have to taper the tail-end of the kinetic energy spectrum.

The behavior of the forecast errors of the HYBRID-1 and HYBRID-2 experiment is notably different from that of the HYBRID-OPT experiment. The values of $\hat{\epsilon}_{max}$ are smaller for these experiments than the HYBRID-OPT experiment (9.57\,m\,s$^{-1}$ and 9.79\,m\,s$^{-1}$ vs. 10.52\,m\,s$^{-1}$). A similar behavior can be observed at the other model levels and other variables (results are not shown for the other variables), which suggests that in the HYBRID-1 and HYBRID-2 experiment, the hybrid model reduces  the root-mean-square error, in part, by reducing the spatial variability of the forecast fields. This is in contrast to the observed increase of the spatial variability of the forecast fields in the HYBRID-OPT experiment compared to the PHYS experiment. Since the only difference between the three hybrid model experiments is the in the training data, this change in the behavior of the hybrid model is caused solely by the different training data. The values of the other parameters, $\alpha$, $\beta$, $c_1$, and $c_2$, are also very different for the HYBRID-1 and HYBRID-2 experiment than for the HYBRID-OPT experiment. Because of the unusual shape of the error growth curves in the HYBRID-1 and HYBRID-2 experiment, we refrain from trying to explain these differences based on the standard interpretation of $\alpha$ and $\beta$. 

\section{Conclusions} \label{sec:conclusions}
In this paper, we investigated the effect of hybridization of a forecast model on the accuracy of the analyses and ensuing forecasts. More specifically, we examined the results of analysis-forecast experiments in which we used a hybridized version of the medium-complexity atmospheric global circulation model SPEEDY as the forecast model. In these experiments, we assimilated simulated observations that were prepared assuming that ERA5 reanalyses represented the true state space trajectory of the atmosphere. These observations were assimilated with a research DA system based on the LETKF DA scheme. Like all other ensemble-based DA scheme, the LETKF uses the forecast model to evolve both the state estimate and the estimate of the uncertainty in the state estimate from one analysis time to the next. Thus, the accuracy of the analyses is a  measure of the quality of the forecast model used in the DA process. The quality of the trained hybrid model was further evaluated by preparing 10-day deterministic forecasts started from the analyses.

The results showed that hybridizing the physics-based model had major analysis and forecast benefits. In terms of analysis accuracy, the benefits were more substantial if the hybrid model was trained on ERA5 reanalyses (HYBRID OPT experiment) rather than analyses obtained with the physics-based model (HYBRID-1 experiment) or the hybrid model trained on analyses obtained with the physics-based model (HYBRID2-experiment). Specifically, the hybridization of the model in the HYBRID-OPT experiment eliminated all but a few highly-localized analysis biases and substantially reduced the magnitude of the transient (flow dependent) analysis errors. The hybrid model was less effective in reducing the analysis biases in the HYBRID-1 and HYBRID-2 experiment. In terms of forecast accuracy, however, the magnitude of the differences between the HYBRID-1, HYBRID-2, and HYBRID-OPT experiment were more modest. In fact, after 1-3 forecast days, the forecast errors were smaller in the HYBRID-1 and HYBRID-2 experiment than the HYBRID-OPT experiment for most variables. This behavior, in part, was the result of a modest decrease of the spatial variability of the forecast fields in the HYBRID-1 and HYBRID-2 experiment. (This can be an undesirable feature in some applications, for example, if the model provides the members of a forecast ensemble, which are expected to capture the full spectrum of forecast uncertainties.) Another likely factor in this behavior was that the training data of the HYBRID-1 and HYBRID-2 experiment were more consistent with the attractor of the hybrid model than the ERA5 reanalyses used for training in the HYBRID-OPT experiment. While it is somewhat disappointing that the results of the HYBRID-2 experiment were not more positive compared to those of the HYBRID-1 experiment, it is possible that they were the results of limitations of the specific implementation of the DA scheme rather than a fundamental limitation of the iterative training approach. For example, using an online bias estimation procedure to better account for the surface pressure background bias in the DA scheme may produce analyses that are better suited for iterative training. 

From the point of view of a potential operational implementation of the investigated hybridization approach, the qualitative differences between the results of our experiments are more relevant than the quantitative differences. On the one hand, a state-of-the-art NWP model would leave less room for improvements by hybridization. On the other hand, the analyses obtained by using the model in DA would produce training data that are more consistent with the model dynamics. Nevertheless, the results suggest that the investigated approach could potentially lead to both analysis and forecast improvements. Compared to other hybridization approaches, it also has the practical advantages that it can be implemented without making changes to the physics-based model and does not require the availability of its linearized version.

\acknowledgments
This research was funded by the Office of Naval Research grant N00014-22-1-2319. Portions of this research were conducted with the advanced computing resources provided by Texas A\&M High Performance Research Computing.
%
%
\datastatement
The code developed and data generated in this research are available at 
\scriptsize{\url{https://github.com/dylanelliotttamu/letkf-hybrid-speedy-grace-training-edition -original}}, \\
 \url{https://zenodo.org/records/16657991}, and \\ 
 \url{https://zenodo.org/records/16659508}.








%




\bibliographystyle{ametsocV6}
\bibliography{references}

\begin{thebibliography}{42}
\providecommand{\natexlab}[1]{#1}
\providecommand{\url}[1]{\texttt{#1}}
\renewcommand{\UrlFont}{\rmfamily}
\providecommand{\urlprefix}{URL }
\expandafter\ifx\csname urlstyle\endcsname\relax
  \providecommand{\doi}[1]{https://doi.org/\discretionary{}{}{}#1}\else
  \providecommand{\doi}{https://doi.org/\discretionary{}{}{}\begingroup
  \urlstyle{rm}\Url}\fi
\providecommand{\eprint}[2][]{\url{#2}}

\bibitem[{Arcomano et~al.(2020)Arcomano, Szunyogh, Pathak, Wikner, Hunt,, and
  Ott}]{Arcomano2020}
Arcomano, T., I.~Szunyogh, J.~Pathak, A.~Wikner, B.~R. Hunt, and E.~Ott, 2020:
  A machine learning-based global atmospheric model. \textit{Geophys.\ Res.\
  Lett.}, \textbf{47}, e2020GL087\,776, \doi{10.1029/2020GL087776}.

\bibitem[{Arcomano et~al.(2023)Arcomano, Szunyogh, Wikner, Hunt,, and
  Ott}]{Arcomano2023}
Arcomano, T., I.~Szunyogh, A.~Wikner, B.~Hunt, and E.~Ott, 2023: A hybrid
  atmospheric model incorporating machine learning can capture dynamical
  processes not captured by its physics-based component. \textit{Geophys.\
  Res.\ Lett.}, \textbf{50}, e2022GL102\,649, \doi{10.1029/2022GL102649}.

\bibitem[{Arcomano et~al.(2022)Arcomano, Szunyogh, Wikner, Pathak, Hunt,, and
  Ott}]{Arcomano2022}
Arcomano, T., I.~Szunyogh, A.~Wikner, J.~Pathak, B.~R. Hunt, and E.~Ott, 2022:
  A hybrid approach to atmospheric modeling that combines machine learning with
  a physics-based numerical model. \textit{J. \ Adv.\ Mod.\ Earth\ Syst.},
  \textbf{14}, e2021MS002\,712, \doi{10.1029/2021MS002712}.

\bibitem[{Baek et~al.(2004)Baek, Hunt, Szunyogh, Zimin,, and Ott}]{Baek2004}
Baek, S.-J., B.~R. Hunt, I.~Szunyogh, A.~Zimin, and E.~Ott, 2004: Localized
  error bursts iin estimating the state of spatiotemporal chaos.
  \textit{Chaos}, \textbf{137}, 2349--2364, \doi{10.1063/1.1788091}.

\bibitem[{Baek et~al.(2009)Baek, Szunyogh, Hunt,, and Ott}]{Baek2009}
Baek, S.-J., I.~Szunyogh, B.~R. Hunt, and E.~Ott, 2009: Correcting for surface
  pressure background bias in ensemble-based analyses. \textit{Mon.\ Wea.\
  Rev.}, \textbf{14}, 1042--1049, \doi{10.1175/2008MWR2787.1}.

\bibitem[{Bi et~al.(2023)Bi, Xie, Zhang, Chen, Gu,, and Tian}]{Bi2023}
Bi, K., L.~Xie, H.~Zhang, X.~Chen, X.~Gu, and Q.~Tian, 2023: Accurate
  medium-range global weather forecasting with 3{D} neural networks.
  \textit{Nature}, \textbf{619}, 533--538, \doi{10.1038/s41586-023-06185-3}.

\bibitem[{Bocquet et~al.(2021)Bocquet, Farchi,, and Malartic}]{Boucquet2021}
Bocquet, M., A.~Farchi, and Q.~Malartic, 2021: Online learning of both state
  and dynamics using ensemble kalman filters. \textit{Foundations Data Sci.},
  \textbf{3}, 305--330, \doi{10.3934/fods.2020015}.

\bibitem[{Brajard et~al.(2020)Brajard, Carassi, Bocquet, Farchi,, and
  Bertino}]{Brajard2020}
Brajard, J., A.~Carassi, M.~Bocquet, A.~Farchi, and L.~Bertino, 2020: Combining
  data assimilation and machine learning to emulate a dynamical model from
  sparse and noisy observations: {A} case study with the {L}orenz 96 model.
  \textit{J.\ Comp. Sci.}, \textbf{44}, 101\,171,
  \doi{10.1016/j.jocs.2020.101171}.

\bibitem[{Dalcher and Kalnay(1987)Dalcher, and Kalnay}]{dalcher1987}
Dalcher, A., and E.~Kalnay, 1987: Error growth and predictability in
  operational {ECMWF} forecasts. \textit{Tellus}, \textbf{39A}, 474--491.

\bibitem[{Farchi et~al.(2021)Farchi, Bocquet, Laloyaux, Bonavita,, and
  Malartic}]{Farchi2021}
Farchi, A., M.~Bocquet, P.~Laloyaux, M.~Bonavita, and Q.~Malartic, 2021: A
  comparison of combined data assimilation and machine learning methods for
  offline and online model error correction. \textit{J.\ Comp. Sci.},
  \textbf{55}, 101\,468, \doi{10.1016/j.jocs.2021.101468}.

\bibitem[{Farchi et~al.(2023)Farchi, Chrust, Bocquet,, and
  Laloyaux}]{Farchi2023}
Farchi, A., M.~Chrust, M.~Bocquet, and P.~Laloyaux, 2023: Online model error
  correction with neural networks in the incremental 4d-var framework.
  \textit{J. \ Adv.\ Mod.\ Earth\ Syst.}, \textbf{15}, e2022MS003\,474,
  \doi{10.1029/2022MS003474}.

\bibitem[{Friedland(1969)}]{Friedland1969}
Friedland, B., 1969: Treatment of bias in recursive filtering. \textit{IEEE
  Trans.\ Autom.\ Contr.}, \textbf{14}, 359--367,
  \doi{10.1109/TAC.1969.1099223}.

\bibitem[{Hatfield(2018)}]{Hatfield2018}
Hatfield, S., 2018: letkf-speedy. github.com,
  \url{https://github.com/samhatfield/letkf-speedy},
  \doi{10.5281/zenodo.1198432}.

\bibitem[{Hersbach et~al.(2020)}]{Herschbach2020}
Hersbach, H., and Coauthors, 2020: The {ERA5} global reanalysis.
  \textit{Quart.\ J.\ Roy.\ Meteor.\ Soc.}, \textbf{146}, 1999--2049,
  \doi{10.1002/qj.3803}.

\bibitem[{Hunt et~al.(2007)Hunt, E.~J,, and Szunyogh}]{Hunt2007}
Hunt, B.~R., K.~E.~J, and I.~Szunyogh, 2007: Efficient data assimilation for
  spatiotemporal chaos: A local ensemble transform kalman filter.
  \textit{Physica D}, \textbf{230}, 112--126,
  \doi{10.1016/j.physd.2006.11.008}.

\bibitem[{Jaeger(2001)}]{Jaeger2001}
Jaeger, H., 2001: The ``echo state'' approach to analysing and training
  recurrent neural networks-with an erratum note. {GMD} {T}echnical {R}eport,
  German Research Center for Information Technology, 148 pp.

\bibitem[{Jazwinski(1970)}]{Jazwinski1970}
Jazwinski, A.~H., 1970: \textit{Stochastic processes and filtering theory}.
  Academic Press, 376 pp.

\bibitem[{Kochkov et~al.(2024)}]{Kochkov2024}
Kochkov, D., and Coauthors, 2024: Neural general circulation models for weather
  and climate. \textit{Nature}, \textbf{632}, 1060--1066,
  \doi{10.1038/s41586-024-07744-y}.

\bibitem[{Kucharski et~al.(2006)Kucharski, Molteni,, and
  Bracco}]{Kucharski2006}
Kucharski, F., F.~Molteni, and A.~Bracco, 2006: Decadal interactions between
  the western tropical {P}acific and north {A}tlantic oscillation.
  \textit{Climate Dyn.}, \textbf{26}, 72--91, \doi{10.1007/s00382-005-0085-5}.

\bibitem[{Kuhl et~al.(2007)}]{Kuhl2007}
Kuhl, D., and Coauthors, 2007: Assessing predictability with a local ensemble
  {K}alman filter. \textit{J.\ Atmos.\ Sci.}, \textbf{64}, 1116--1140,
  \doi{doi.org/10.1175/JAS3885.1}.

\bibitem[{Laloyaux et~al.(2020)Laloyaux, Bonavita, Dahoui, Farnan, Healy,
  H'{o}lm,, and Lang}]{Laloyaux2020}
Laloyaux, P., M.~Bonavita, M.~Dahoui, J.~Farnan, S.~Healy, E.~H'{o}lm, and
  S.~T.~K. Lang, 2020: Towards an unbiased stratospheric analysis.
  \textit{Quart.\ J.\ Roy.\ Meteor.\ Soc.}, \textbf{146}, 2392--2409,
  \doi{10.1002/qj.3798}.

\bibitem[{Lam et~al.(2023)}]{Lam2023}
Lam, R., and Coauthors, 2023: Learning skillful medium-range global weather
  forecasting. \textit{Science}, \textbf{382}, 1416--1421,
  \doi{10.1126/science.adi2336}.

\bibitem[{Lorenz(1982)}]{Lorenz1982}
Lorenz, E.~N., 1982: Atmospheric predictability experiments with a large
  numerical model. \textit{Tellus}, \textbf{34A}, 505--513.

\bibitem[{Luko\v{s}evi\v{c}ius(2012)}]{Lukosevicius2012}
Luko\v{s}evi\v{c}ius, M., 2012: A practical guide to applying echo state
  networks. \textit{{Neural networks: Tricks of the trade 2nd edition}},
  G.~Montavon, G.~B. Orr, and K.-R. M\"{u}ller, Eds., Springer, 659--686.

\bibitem[{Luko\v{s}evi\v{c}ius and Jaeger(2009)Luko\v{s}evi\v{c}ius, and
  Jaeger}]{Lukosevicius2009}
Luko\v{s}evi\v{c}ius, M., and H.~Jaeger, 2009: Reservoir computing approaches
  to recurrent neural network training. \textit{Computer Science Review},
  \textbf{3}, 127--149.

\bibitem[{Malartic et~al.(2022)Malartic, Farchi,, and Bocquet}]{Malartic2022}
Malartic, Q., A.~Farchi, and M.~Bocquet, 2022: State, global, and local
  parameter estimation using local ensemble kalman filters: Applications to
  online machine learning of chaotic ddynamics. \textit{Quart.\ J.\ Roy.\
  Meteor.\ Soc.}, \textbf{148}, qj.4297--2193, \doi{10.1002/qj.4297}.

\bibitem[{Molteni(2003)}]{Molteni2003}
Molteni, F., 2003: Atmospheric simulations using a gcm with simplified physical
  parameterizations. i: Model climatoology and variability in multi-decadal
  experiments. \textit{Climate Dyn.}, \textbf{20}, 175--191,
  \doi{10.1007/s00382-002-0268-2}.

\bibitem[{Ott et~al.(2004)}]{Ott2004}
Ott, E., and Coauthors, 2004: A local ensemble kalman filter for atmospheric
  data assimilation. \textit{Tellus}, \textbf{56A}, 415--428,
  \doi{10.1111/j.1600-0870.2004.00076.x}.

\bibitem[{Park et~al.(2023)Park, Xue,, and Liu}]{Park2023}
Park, J., M.~Xue, and C.~Liu, 2023: Implementation and testing of radar data
  assimilation capabilities within the {J}oint {E}ffort for {D}ata assimilation
  {I}ntegration framework with ensemble transformation {K}alman filter coupled
  with {FV3}-{LAM} model. \textit{Geophys.\ Res.\ Lett.}, \textbf{50},
  e2022GL102\,709, \doi{10.1029/2022GL102709}.

\bibitem[{Patel et~al.(2024)Patel, Arcomano, Hunt, Szunyogh,, and
  Ott}]{Patel2023}
Patel, D., T.~Arcomano, B.~R. Hunt, I.~Szunyogh, and E.~Ott, 2024: Exploring
  the potential of hybrid machine-learning/physics-based modeling for
  atmospheric/oceanic prediction beyond the medium range. \textit{J. \ Adv.\
  Mod.\ Earth\ Syst.}

\bibitem[{Pathak et~al.(2018)Pathak, Wikner, Fussel, Chandra, Hunt, Girvan,,
  and Ott}]{Pathak2018}
Pathak, J., A.~Wikner, R.~Fussel, S.~Chandra, B.~R. Hunt, M.~Girvan, and
  E.~Ott, 2018: Hybrid forecasting of chaotic processes: using machine learning
  in conjunction with a knowledge-based model. \textit{Chaos}, \textbf{28},
  041\,101, \doi{10.1063/1.5028373}.

\bibitem[{Pathak et~al.(2022)}]{Pathak2022}
Pathak, J., and Coauthors, 2022: Fourcastnet: a global data-driven
  high-resolution weather model using adaptive fourier neural operators.
  \textit{arXiv}, \doi{10.48550/arXiv.2202.11214}.

\bibitem[{Szunyogh(2014)}]{Szunyogh2014}
Szunyogh, I., 2014: \textit{Applicable atmospheric dynamics: Techniques for the
  exploration of atmospheric dynamics}. World Scientific, 588 pp.,
  \doi{10.1142/8047}.

\bibitem[{Szunyogh et~al.(2008)Szunyogh, Kostelich, Gyarmati, Kalnay, Hunt,
  Ott, Satterfieldl,, and Yorke}]{Szunyogh2008}
Szunyogh, I., E.~J. Kostelich, G.~Gyarmati, E.~Kalnay, B.~R. Hunt, E.~Ott,
  E.~Satterfieldl, and J.~A. Yorke, 2008: A local ensemble transform {K}alman
  filter data assimilation system for the {NCEP} global model. \textit{Tellus},
  \textbf{60A}, 113--130, \doi{10.1111/j.1600-0870.2007.00274.x}.

\bibitem[{Szunyogh et~al.(2005)Szunyogh, Kostelich, Gyarmati, Patil, Hunt,
  Kalnay, Ott,, and Yorke}]{Szunyogh2005}
Szunyogh, I., E.~J. Kostelich, G.~Gyarmati, D.~J. Patil, B.~R. Hunt, E.~Kalnay,
  E.~Ott, and J.~A. Yorke, 2005: Assessing a local ensemble kalman filter:
  perfect model experiments with the {N}ational {C}enters for {E}nvironmental
  {P}rediction global model. \textit{Tellus}, \textbf{57A}, 528--545,
  \doi{10.3402/tellusa.v57i4.14721}.

\bibitem[{Tikhonov and Arsenin(1977)Tikhonov, and
  Arsenin}]{Tikhonov+Arsenin1977}
Tikhonov, A.~N., and V.~V. Arsenin, 1977: \textit{Solutions of ill-posed
  problems}. Winston \& {S}ons, 272 pp.

\bibitem[{Tr\'{e}molet(2006)}]{Tremolet2006}
Tr\'{e}molet, Y., 2006: Accounting for an imperfect model in 4{D}-{V}ar.
  \textit{Quart.\ J.\ Roy.\ Meteor.\ Soc.}, \textbf{132}, 2483--2504,
  \doi{10.1256/qj.05.224}.

\bibitem[{\v{Z}agar et~al.(2017)\v{Z}agar, Horvath, \v{Z}iga Zaplotnik,, and
  Magnusson}]{zagar2017}
\v{Z}agar, N., M.~Horvath, \v{Z}iga Zaplotnik, and L.~Magnusson, 2017:
  Scale-dependent estimates of the growth of forecast uncertainties in a global
  prediction system. \textit{Tellus}, \textbf{69}, 1287--492.

\bibitem[{Weyn et~al.(2021)Weyn, Durran, Caruana,, and
  Cresswell-Clay}]{Weyn2021}
Weyn, J.~A., D.~R. Durran, R.~Caruana, and N.~Cresswell-Clay, 2021:
  Sub-seasonal forecasting with a large ensemble of deep-learning weather
  prediction models. \textit{J. \ Adv.\ Mod.\ Earth\ Syst.}, \textbf{13},
  e2021MS002\,502, \doi{10.1029/2021MS002502}.

\bibitem[{Wikner et~al.(2024)Wikner, Harvey, Girvan, Hunt, Pomerance,
  Antonsen,, and Ott}]{Wikner2024}
Wikner, A., J.~Harvey, M.~Girvan, B.~R. Hunt, A.~Pomerance, T.~Antonsen, and
  E.~Ott, 2024: Stabilizing machine learning prediction of dynamics: Novel
  noise-inspired regularization tested with reservoir computing. \textit{Neural
  Networks}, \textbf{170}, 94--110, \doi{0.1016/j.neunet.2023.10.054}.

\bibitem[{Wikner et~al.(2020)Wikner, Pathak, Hunt, Girvan, Arcomano, Szunyogh,
  Pomerance,, and Ott}]{Wikner2020}
Wikner, A., J.~Pathak, B.~R. Hunt, M.~Girvan, T.~Arcomano, I.~Szunyogh,
  A.~Pomerance, and E.~Ott, 2020: Combining machine learning with
  knowledge-based modeling for scalable forecasting and subgrid-scale closure
  of large, complex, spatiotemporal systems. \textit{Chaos}, \textbf{30},
  053\,111, \doi{10.1063/5.0005541}.

\bibitem[{Wikner et~al.(2021)Wikner, Pathak, Hunt, Szunyogh, Girvan,, and
  Ott}]{Wikner2021}
Wikner, A., J.~Pathak, B.~R. Hunt, I.~Szunyogh, M.~Girvan, and E.~Ott, 2021:
  Using data assimilation to train a hybrid forecast system that combines
  machine-learning and knowledge-based components. \textit{Chaos}, \textbf{31},
  053\,114, \doi{10.1063/5.0048050}.

\end{thebibliography}

\end{document}